\documentclass[11pt,a4paper]{article}
\usepackage[utf8]{inputenc}
\usepackage[T1]{fontenc}
\usepackage{lmodern}
\usepackage[dvipsnames]{xcolor}
\usepackage{graphicx}
\usepackage{amsmath,amssymb,amsfonts,mathtools}
\usepackage{bm}
\usepackage[font=small]{caption}
\usepackage{subcaption}
\usepackage{booktabs}
\usepackage{multirow}
\usepackage{array}
\usepackage{tabularx}
\usepackage{float}
\usepackage{placeins}
\usepackage{algorithm}
\usepackage{algpseudocode}
\usepackage{hyperref}
\usepackage{cleveref}
\usepackage{ragged2e}
\usepackage{siunitx}
\usepackage{enumitem}
\usepackage{nicefrac}
\usepackage{adjustbox}
\usepackage{url}
\usepackage[numbers,sort&compress]{natbib}
\usepackage{authblk}
\usepackage[margin=1in]{geometry}
\usepackage{xspace}
\graphicspath{{./}}

\hypersetup{
  colorlinks = true,
  urlcolor   = blue,
  linkcolor  = blue,
  citecolor  = blue,
  pdfauthor  = {Shaoliang Yang, Jun Wang, Yunsheng Wang},
  pdftitle   = {IterSIMP-sigma: Evaluating LLM-Assisted Spatial Interventions in Stress-Aware Topology Optimization},
  pdfkeywords = {design automation, topology optimization, large language models in design, stress-aware design, SIMP},
  pdfsubject = {Comprehensive arXiv preprint for the journal-aligned study}
}


\newcommand{\sigyield}{\sigma_{\mathrm{yield}}}
\newcommand{\sigvm}{\sigma_{\mathrm{vm}}}
\newcommand{\rhobar}{\bar{\rho}}

\newcommand{\Crep}{C_{\mathrm{rep}}}
\newcommand{\Cfinal}{C_{\mathrm{final}}}
\newcommand{\methodname}{IterSIMP-$\sigma$}

\newcolumntype{Y}{>{\RaggedRight\arraybackslash}X}

\begin{document}

\title{IterSIMP-$\sigma$: Evaluating LLM-Assisted Spatial Interventions in Stress-Aware Topology Optimization}

\author[1]{Shaoliang Yang}
\author[1]{Jun Wang\thanks{Corresponding author. E-mail: jwang22@scu.edu}}
\author[1]{Yunsheng Wang}
\affil[1]{Department of Mechanical Engineering, Santa Clara University, Santa Clara, CA 95053, USA}

\date{}
\maketitle

\begin{abstract}
This paper studies whether multimodal large language models (LLMs) can serve
as inspectable spatial proposal modules for stress-aware topology optimization.
IterSIMP-$\sigma$ keeps the SIMP optimizer as a compliance-minimizing
finite-element solver and places a deterministic stress pass, gate evaluator,
and hybrid LLM/rule interpreter around it.  After each solve, density and von
Mises stress fields are rendered; the interpreter proposes ranked spatial
interventions; and deterministic safeguards accept, reject, or stop each
action.  The main action is a soft density seed, where selected elements are
initialized at elevated density before the next solve but remain free under the
optimality-criteria update.  We evaluate the loop on a 16-problem 2D
controller-policy benchmark, a six-problem exploratory 3D extension,
passive-solid and input ablations, stress-threshold sensitivity, and a
fixed-volume attribution study comparing LLM proposals with deterministic
max-stress hotspot seeding, random stress-region seeding, and rule-based
control.  The 2D controller-policy benchmark shows a small retained-compliance
difference (1.9\% lower geometric mean for the soft-seed LLM), but this
diagnostic is not statistically significant ($W=33$, two-sided $p=0.382$)
and is not a fixed-volume feasible-final comparison.  In the
fixed-volume study, the LLM condition completed 44/48 attempted evaluations;
25/44 completed evaluations produced all-gate-passing retained states.
Feasible-final scoring against rule-based control is split 4/4/1, and
deterministic exact-hotspot seeding remains competitive.  Accepted LLM spatial
actions with per-step records have mean normalized seed-to-hotspot distance
0.221.  The results support IterSIMP-$\sigma$ as an inspectable LLM-assisted
design-automation framework for spatial interventions, not yet as evidence that
LLM visual reasoning improves stress-constrained optimization.
\end{abstract}


\section{Introduction}
\label{sec:intro}

Topology optimization is a core computational design methodology for
distributing material within a prescribed design domain to improve structural
performance under volume, manufacturing, and increasingly strength-related
requirements~\cite{bendsoe2003topology,sigmund2013topology}.  Early
homogenization work helped establish material-layout optimization as a
structural-design tool~\cite{bendsoe1988generating}.  The Solid Isotropic
Material with Penalization (SIMP) method remains one of the most widely used
density-based formulations because it combines finite-element analysis,
penalized stiffness interpolation, filtering, projection, and
optimality-criteria updates in a computationally efficient design
loop~\cite{bendsoe2003topology,sigmund2001design,andreassen2011efficient,wang2011projection,lazarov2011filters}.
Even in compliance-only settings, the final topology depends strongly on
continuation choices such as penalization, filter radius, projection
sharpness, move limits, and stopping criteria.  These choices are often treated
as numerical details, but in practice they shape the designer's ability to
obtain crisp, feasible, and mechanically interpretable
layouts~\cite{sigmund1998numerical}.

Stress-aware topology optimization makes the design task more difficult.
Element-wise stress constraints are numerous, nonlinear, and sensitive to the
stress singularity associated with vanishing material density
regions~\cite{duysinx1998topology,le2010stress}.  Classical density-based
approaches therefore aggregate local stress measures into a smaller set of
constraints, commonly through $P$-norm or Kreisselmeier--Steinhauser
functions~\cite{yang1996stress,le2010stress,holmberg2013stress}.  More recent
augmented-Lagrangian and local-constraint methods reduce reliance on
aggregation and have enabled large two- and three-dimensional demonstrations,
but local stress control remains computationally demanding and sensitive to
modeling choices~\cite{senhora2020topology,dasilva2021threedim,senhora2023local,da2025stress,pmc2024review}.
The design consequence is visible: stress-limited layouts often round
re-entrant corners, widen thin members, and redistribute load paths relative to
compliance-optimal layouts~\cite{bruggi2012stress,giraldolondono2020polystress}.

This paper approaches that visual and geometric character as a design
automation opportunity.  A structural designer inspecting a density field and a
von Mises stress field can often identify a localized stress concentration,
infer whether it arises from a sharp corner, under-dimensioned member, or
asymmetric load path, and propose a targeted geometric modification.  Standard
continuation schedules and scalar rule-based controllers do not provide that
spatial vocabulary.  They can change global parameters, but they cannot
directly express interventions such as reinforcing a hotspot, widening a
stressed member, or redistributing material between low- and high-stress
regions.

Machine-learning methods provide one route to using richer field information
in topology optimization.  Conditional generators and convolutional networks
have been trained to predict final topologies from loads, boundary conditions,
intermediate density fields, or initial finite-element
fields~\cite{yu2019deep,sosnovik2019neural,nie2021topologygan}.  Other
approaches embed learned models inside the optimization loop to accelerate
convergence or guide intermediate density
trajectories~\cite{wang2022selfdirected,rade2021algorithmically,jcde2023review}.
These methods show that field-aware learning can improve design workflows, but
they usually require task-specific training data and do not expose a
human-readable mechanism for proposing local engineering interventions during a
solve.  Warm-starting methods similarly demonstrate the practical importance of
initial density fields, but typically rely on learned priors or precomputed
datasets~\cite{zhang2024degm,white2024conditioning,kallioras2020accelerated}.

We study a narrower and more inspectable role for multimodal large language
models (LLMs): using visual reasoning as an outer-loop spatial proposal module
around an existing SIMP solver.  Recent multimodal models can interpret images
and diagrams in addition to text~\cite{achiam2023gpt,team2023gemini}, and
LLM-based agents are increasingly used to coordinate engineering design,
structural analysis, and optimization
workflows~\cite{makatura2024llm,tian2024shearwall,llm_structural_analysis2025,orllm2025,llm_meta_optimizers2026}.
For stress-aware topology optimization, the relevant capability is not
open-ended design generation.  The relevant capability is bounded spatial
reasoning: reading density and stress-field images, identifying a plausible
local design weakness, and returning a structured intervention that a
deterministic optimizer can accept, reject, or evaluate.

\subsection{LLM vision as spatial reasoning for topology optimization}

The technical question motivating this work is whether an LLM can
(i) localize stress concentrations from a rendered von Mises stress field,
(ii) relate those concentrations to the current density topology,
(iii) propose geometry modifications with usable spatial coordinates, and
(iv) express those modifications in a form that the SIMP optimizer can exploit
without surrendering its gradient-based degrees of freedom.

In prior work, we studied an LLM-based adaptive continuation controller for
compliance-only SIMP~\cite{yang2026itersimp}, and a natural-language
topology-optimization pipeline that can configure, solve, and validate
structural problems~\cite{yang2026autosimp}.  The compliance-controller setting
used a single density-field image per design step and did not perform
stress-localized geometry intervention.  Stress-aware spatial
control requires a different capability: the controller must read two fields
simultaneously, identify the spatial location of stress violations, and propose
geometry changes that are precise enough to resolve local concentrations
without disrupting global load paths.

\subsection{Contributions}

This paper makes the following contributions for design automation and
topology optimization:

\begin{enumerate}

\item[(1)] \textbf{Dual-image LLM interpreter for stress-aware SIMP.}
We introduce a stress-aware LLM interpreter that accepts two simultaneous visual
inputs per design step: the element density field and the per-element von Mises
stress field.  The controller can therefore use observed stress concentrations,
stress concentration factors, and constraint-violation status when proposing
geometry modifications.  We add three stress-aware spatial action types:
hotspot reinforcement, stressed-member widening, and material redistribution.

\item[(2)] \textbf{Controlled soft-seed/passive-solid materialization study.}
We compare a seed-biased initialization strategy in which LLM-proposed regions
are initialized at elevated density but remain subject to the optimality-criteria update
against a passive-solid variant that freezes LLM-placed material.  The
three-condition ablation (Section~\ref{sec:ablation}) makes the central
mechanistic tradeoff explicit: soft seeding is less restrictive, while
passive-solid materialization is the stronger aggregate performer in the
reported retained-compliance data.

\item[(3)] \textbf{Fixed-volume attribution study with deterministic and
input-ablation baselines.}
We define a 16-problem two-dimensional stress-gated controller-policy benchmark
plus a fixed-volume attribution study that holds the initial volume fraction constant and
compares the LLM policy with deterministic max-stress hotspot seeding, random
stress-region seeding, rule-based control, and density-only, stress-only,
numeric-only, and global-action-only LLM ablations.  This is designed to probe
whether any apparent LLM advantage persists after material-budget changes are
disabled.

\item[(4)] \textbf{Direct localization metrics for spatial actions.}
For the fixed-volume attribution study, where per-step traces are available, we measure the
distance between accepted spatial seed centers and the true numerical
max-stress element, overlap with the top 1\%, 5\%, and 10\% stress regions, and
next-step stress changes where the trace supports that coupling.  These metrics
test whether LLM proposals target stress hotspots rather than treating visual
examples as qualitative evidence.

\item[(5)] \textbf{Benchmark, reproducibility map, and failure-mode analysis.}
We report the original 16-problem two-dimensional controller-policy
comparison, a six-problem exploratory three-dimensional extension, and a
fixed-volume attribution study with explicit source mapping.  The benchmark
covers cantilevers, bridges, brackets, portal frames, Michell trusses,
asymmetric loads, and torsion-loaded 3D domains.  Because the LLM uses
deterministic decoding, repeat seeds identify solver and controller traces
rather than independent LLM samples.  We report descriptive per-problem
statistics, feasibility-conditioned endpoints, two-sided Wilcoxon tests,
geometric-mean ratios, final gate-pass counts, seed-level variance diagnostics,
and a validity check for the 3D bridge result.

\end{enumerate}

\subsection{Scope and limitations}

The present study uses Gemini 3.1 Flash-Lite Preview as the single LLM at
temperature $T=0$ (deterministic decoding). The corresponding model identifier
is \texttt{gemini-\allowbreak{}3.1-\allowbreak{}flash-\allowbreak{}lite-\allowbreak{}preview}~\cite{google2026gemini31flashlite}.
The experiments used the preview endpoint available at execution time, and no
served-model revision metadata was retained.
The repeat seeds are recorded
with each run; in the current implementation, deterministic solver paths begin
from a uniform volume-fraction density, so seed-level variation is limited to
paths where the controller or solver state differs across repeats.  Archived
records support code-level replay of solver and controller state, but
not exact replay of the original hosted-model calls.  The 3D
benchmark provides an exploratory three-seed controller comparison against a
rule-based baseline (Section~\ref{subsec:3d}).  This comparison should be
interpreted as policy-level evidence rather than an isolated estimate of the
seed mechanism: the reported 3D subset does not include a no-seed 3D ablation,
fixed-volume attribution studies, or separately calibrated 3D stress thresholds.

The sensitivity of results to the $\sigyield$ threshold is characterized across
six percentile levels, but repeated runs are reported only at the 50th
percentile.  The fixed-volume 2D attribution study adds deterministic numerical
hotspot seeding, random stress-region seeding, density-only, stress-only,
numeric-only, and global-action-only LLM ablations, plus direct
hotspot-localization measurements.  These runs are the most direct attribution
diagnostic because they hold volume fraction fixed and score feasible-retained,
feasible-final, and retained-any endpoints separately while reporting execution
completeness separately from design feasibility.  They are not pooled with the
original controller-policy comparison because they use a separately calibrated
$\sigyield=163.457$ stress threshold rather than the original
$\sigyield=120.3$ threshold; the former was recomputed on fixed-volume
attribution reruns under the volume-lock protocol, whereas the latter comes
from the original controller-policy trace population.  The controller-policy result should therefore be read as
a benchmark diagnostic, while the fixed-volume study is the
more controlled test of whether LLM spatial proposals outperform deterministic
stress-field baselines.  The manuscript still does not include a $P$-norm,
K--S, or augmented-Lagrangian stress-constrained solver baseline, fixed-volume
3D reruns, or exact hosted-model replay metadata.

\subsection{Paper organization}

Section~\ref{sec:related} reviews related work.
Section~\ref{sec:method} describes the \methodname{} architecture.
Section~\ref{sec:experiments} defines the benchmark and protocol.
Section~\ref{sec:results} reports the main controller-policy results,
beginning with the 2D benchmark comparison and followed by the fixed-volume
attribution study, sensitivity analysis, convergence behavior, and the 3D
benchmark.  Section~\ref{sec:ablation} presents the ablation study.
Section~\ref{sec:discussion} discusses the operating window and failure modes.
Section~\ref{sec:conclusion} concludes.

\section{Related Work}
\label{sec:related}

\subsection{Topology optimization under stress constraints}

Stress-constrained topology optimization has a long and technically rich
history rooted in the observation that compliance-minimizing designs
routinely violate local strength criteria.
Duysinx and Bends{\o}e~\cite{duysinx1998topology} introduced $\varepsilon$-relaxation
of local von~Mises constraints into the SIMP framework, enabling
gradient-based solvers to navigate the singularity problem---the
degenerate sub-spaces of the feasible design domain in which globally
optimal solutions reside but nonlinear programming algorithms cannot
converge.
Yang and Chen~\cite{yang1996stress} and Le et al.~\cite{le2010stress}
established the $P$-norm and Kreisselmeier--Steinhauser (K--S) aggregation
approaches, which compress the exponentially large set of local constraints
into a single global scalar, making the problem tractable at the cost of
conservatism: the aggregation bound is always weaker than the element-wise
bound it approximates.

Holmberg et al.~\cite{holmberg2013stress} addressed the gap between
aggregated and local stress control through element clustering, partitioning
the domain into groups of stress evaluation points and applying a modified
$P$-norm within each cluster.
Bruggi and Duysinx~\cite{bruggi2012stress} studied the interplay between
compliance and stress objectives, showing that minimum-weight designs under
stress constraints differ qualitatively from compliance-optimal topologies.
Giraldo-Londo\~{n}o and Paulino~\cite{giraldolondono2020polystress}
unified the treatment of multiple failure criteria (von~Mises,
Drucker--Prager, Tresca) within a single topology optimization framework.
Applied to the L-bracket and MBB-beam benchmarks---both of which appear in
our evaluation---they demonstrated that stress-constrained formulations
produce qualitatively different topologies from compliance-only designs,
with re-entrant corners replaced by smooth curves.
This geometric observation is central to our motivation: the topological
difference induced by stress constraints is visually legible, making it
a natural target for visual interpretation.

The aggregation-dominated literature was challenged by
Senhora et al.~\cite{senhora2020topology}, who reformulated
stress-constrained topology optimization as a mass-minimization problem via the augmented
Lagrangian method, preserving the local nature of stress constraints without
aggregation.
Their approach requires only one adjoint vector regardless of the number of
constraints, and produces designs that are consistent under mesh refinement.
Xu et al.~\cite{xu2021stress} extended this line to multi-material topology optimization using
ordered SIMP interpolation with a stability transformation method to close
the gap between the $P$-norm stress and the exact maximum local stress.
Da Silva and Emmendoerfer Jr.~\cite{da2025stress} recently conducted a systematic
comparison of augmented Lagrangian subproblem solvers for stress-constrained
topology optimization, confirming the method's scalability to large 3D problems.
A 2024 review~\cite{pmc2024review} surveys the state of the field,
noting that local stress control and convergence robustness remain open
challenges for density-based methods.
Da Silva et al.~\cite{dasilva2021threedim} demonstrated stress-constrained
topology optimization at industrial scale with hundreds of millions of local constraints,
while Granlund et al.~\cite{granlund2023nonproportional} extended
stress-constrained formulations to nonproportional loading.
Da Silva and Emmendoerfer~\cite{dasilva2024failsafe} recently addressed
fail-safe stress-constrained design under manufacturing errors.
Senhora et al.~\cite{senhora2023local} further advanced the augmented
Lagrangian approach to handle continuously varying load directions with
local stress constraints.

Beyond density methods, level-set approaches have also been applied to
stress-constrained topology optimization.
Allaire and Jouve~\cite{allaire2008minimum} demonstrated minimum-stress design
using shape and topological derivatives within a level-set framework.
Guo et al.~\cite{guo2011stress} developed a stress-related level-set formulation
published in this journal, and Picelli et al.~\cite{picelli2018stress}
extended the approach to handle $P$-norm stress aggregation with adaptive
constraint scaling.
Emmendoerfer and Fancello~\cite{emmendoerfer2016level} proposed a
reaction--diffusion-based level-set evolution for local stress constraints.
Xia et al.~\cite{xia2018stress} addressed stress-based topology optimization using the
bi-directional evolutionary structural optimization (BESO) method,
demonstrating that discrete design representations avoid the stress
singularity problem inherent in intermediate-density formulations.
Unlike level-set or BESO formulations, \methodname{} remains density-based:
the stress pass is recovered from the penalized SIMP density field and the
gate is applied over thresholded solid elements.  This makes the stress field
operationally useful for gating and visualization, but it is not equivalent to
boundary-based stress evaluation.

The present work does not embed stress aggregation in the inner optimizer
itself---we retain a post-solve maximum-stress gate over solid elements---but instead uses the LLM
to identify and act on stress concentration locations before the optimizer
converges, reducing the landscape difficulty that aggregation-based methods
face.

\subsection{Machine learning and deep learning for topology optimization}

The application of neural networks to topology optimization has proceeded
along two distinct axes: \emph{data-driven prediction}, which learns a
mapping from boundary conditions to optimal topologies using pre-solved
training sets, and \emph{optimization acceleration}, which embeds neural
networks into the iterative optimization loop itself.

In the prediction line, Yu et al.~\cite{yu2019deep} used conditional
generative adversarial networks trained on 100,000 compliance-minimized
solutions to predict near-optimal 2D topologies without any solver
iterations, demonstrating strong generalization to unseen loads and boundary
conditions.
Nie et al.~\cite{nie2021topologygan} introduced TopologyGAN, conditioning
the generator not just on boundary conditions but also on the von~Mises
stress and strain energy density fields from an initial FEA pass, which is
conceptually related to our dual-image approach---though TopologyGAN
uses field images as \emph{training input features} rather than as
\emph{runtime visual queries} to an agent.
Sosnovik and Oseledets~\cite{sosnovik2019neural} trained convolutional
encoder-decoder networks to predict final topologies from partially
converged intermediate solutions, reducing the number of SIMP iterations
required to reach a specified quality level.

In the acceleration line, Deng et al.~\cite{wang2022selfdirected}
introduced SOLO, which embeds a deep neural network inside the optimization
loop as a dynamically adapting surrogate for the objective function,
updating the model only in the region of interest identified by current
predictions.
Rade et al.~\cite{rade2021algorithmically} developed density sequence (DS)
and coupled density--compliance sequence (CDCS) prediction networks that
enforce physical consistency by training on intermediate iteration
trajectories rather than final solutions only.
A 2023 review by Shin et al.~\cite{jcde2023review} provides comprehensive
coverage of ML-based topology optimization methods, distinguishing iterative from
non-iterative approaches and single-stage from multi-stage architectures.

\methodname{} occupies a distinct position from all of these methods.
It does not predict topologies from training data, nor does it substitute
a neural surrogate for the FEA computation.
Instead, it acts as an \emph{agentic controller} that guides an unmodified
SIMP solver through a sequence of geometry modifications---retaining the
existing SIMP update structure while adding an outer-loop controller that
uses rendered field information.

\subsection{Warm-starting and density initialization in topology optimization}

The sensitivity of SIMP to density initialization is well established.
Uniform initialization ($\rho_e = v_f$) is standard but can trap the
solver in suboptimal local minima, particularly under stress constraints
where the loss landscape is non-convex and multi-modal.

Data-driven warm-starting has emerged as one strategy to improve
initialization quality.
Zhang et al.~\cite{zhang2024degm} trained a deep generative model on
topology optimization solutions and used its predictions as initial density
fields, reducing convergence time by 36--58\% compared to uniform
initialization.
Chen et al.~\cite{white2024conditioning} proposed conditioning field
initialization for neural network-based topology optimization: supplying a strain energy
density field computed on the initial domain as an additional input to the
density network, improving convergence on problems with complex loading.
Kallioras et al.~\cite{kallioras2020accelerated} combined SIMP with
long short-term memory (LSTM) networks to predict and accelerate the
evolution of the density distribution across iterations.

The soft density seed mechanism in \methodname{} differs from these
approaches in an important respect: it does not rely on a pre-trained
model or a learned prior.
Instead, it uses the LLM's runtime spatial targeting to select seed
locations problem-by-problem, warm-starting individual elements at
elevated density based on visual identification of stress concentration
sites.
The mechanism is conceptually closer to \emph{problem-specific
initialization guidance} than to data-driven warm-starting, and its
contribution is quantified by the three-condition ablation in
Section~\ref{sec:ablation}.

\subsection{LLM agents in engineering design and optimization}

Large language models have recently demonstrated broad applicability in
engineering design contexts.
Makatura et al.~\cite{makatura2024llm} conducted a comprehensive survey
of LLM-enabled computational design and manufacturing pipelines, showing
that general-purpose models (GPT-4) can coordinate existing solvers,
algorithms, and visualizers into integrated design workflows without
domain-specific fine-tuning.
They identified \emph{inverse design} and \emph{topology optimization}
as primary candidate domains for LLM augmentation, noting that both
require modularisation of hard problems into tractable subproblems---a
structure amenable to LLM-guided iteration.

Qin et al.~\cite{tian2024shearwall} proposed an LLM-based system
for shear wall structural design, in which the LLM translates engineer
language descriptions into executable optimization code and coordinates
a three-level, two-stage topology--pattern--size pipeline.
This demonstrates LLM-as-controller for structural optimization but does
not involve visual interpretation of field quantities.
For automated structural analysis, recent work~\cite{llm_structural_analysis2025}
used LLMs to parse structural descriptions and generate executable
finite-element scripts, with GPT-4o and Gemini Pro as back-end models.
In broader optimization contexts, the OR-LLM-Agent framework
\cite{orllm2025} applied reasoning LLMs to operations research problems
through a three-stage pipeline (modelling, code generation, debugging)
and showed that structured multi-agent decomposition outperforms
single-prompt approaches on complex optimization instances.
LLM-driven meta-optimization surveys~\cite{llm_meta_optimizers2026}
document the emerging paradigm of LLMs generating, adapting, and
evaluating optimization algorithms rather than serving purely as
natural-language interfaces.

\methodname{} provides an early controlled evaluation of LLM vision over
finite-element density and stress fields as the primary input to a
topology-optimization control loop.
Our prior work studied an LLM-based adaptive continuation controller for
compliance-only SIMP~\cite{yang2026itersimp}, and a
natural-language-to-topology pipeline that can autonomously configure, solve,
and validate structural problems~\cite{yang2026autosimp}.
The present work builds on that LLM-controlled SIMP direction by introducing
dual-image (density and stress) visual reasoning and a soft density seed
mechanism for stress-gated spatial intervention.
Whereas existing systems use LLMs to generate code, parameterize
constraints, or translate natural-language problem descriptions, we
use the LLM's multimodal vision capability to localize stress concentrations
in rendered density and stress field images and propose spatially
targeted geometry modifications.
The empirical evidence distinguishes two effects: the results are consistent
with possible benefits from LLM spatial localization on selected cases, while
the aggregate soft-seed evidence is mixed and not statistically significant.

\section{Method}
\label{sec:method}

\subsection{System overview}

\begin{figure}[H]
  \centering
  \includegraphics[width=\linewidth]{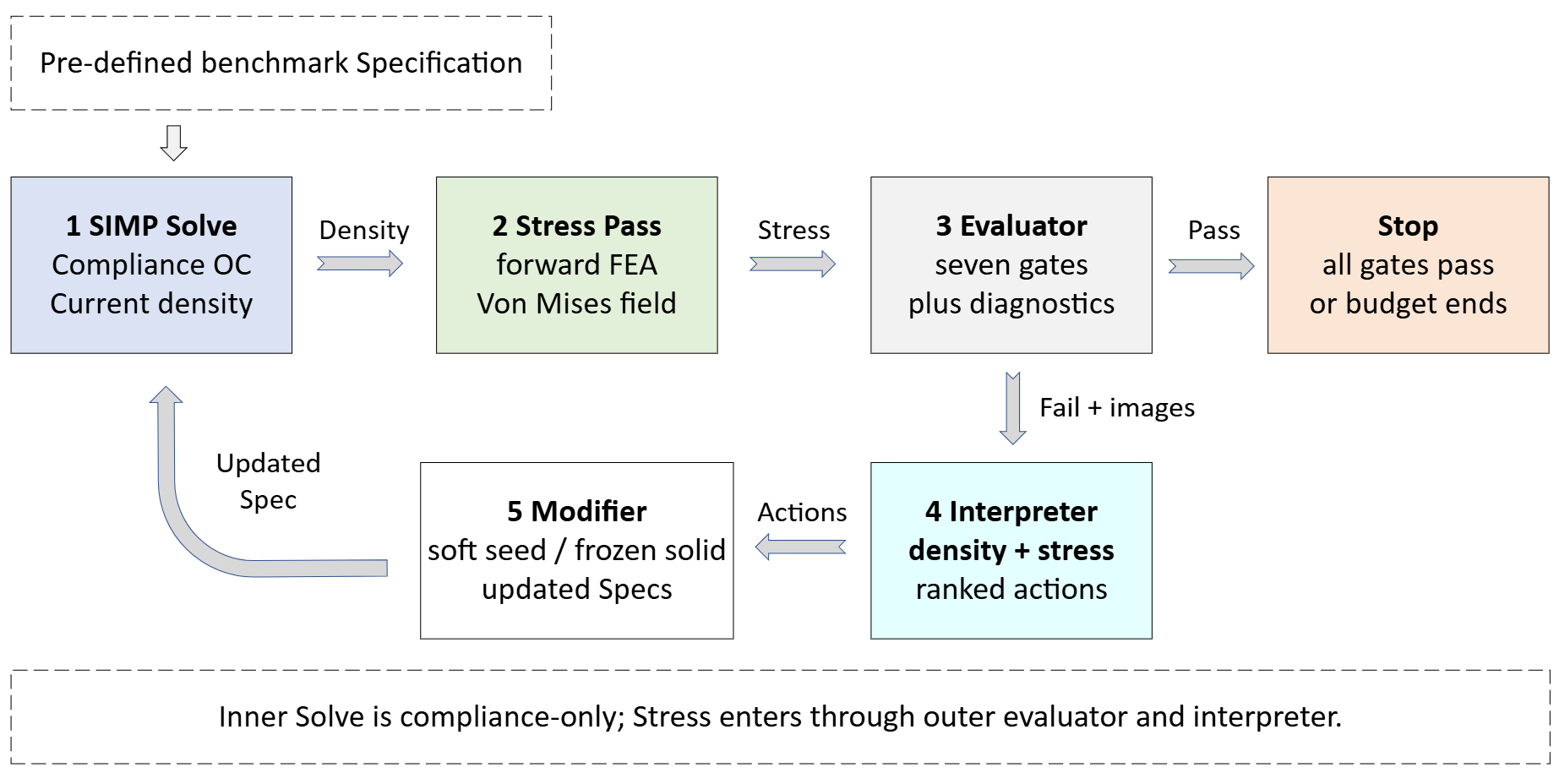}
  \caption{Schematic overview of the benchmark \methodname{} loop.
    The inner SIMP solve remains a compliance-only OC update.  A separate
    stress pass computes the von~Mises field, the evaluator applies seven gate
    checks plus diagnostics, and the LLM/rule interpreter receives the density
    and stress images with gate status to propose ranked actions.  The modifier
    applies the first admissible action as a soft seed, frozen solid, or other
    specification update before returning to the SIMP solve.  The loop stops
    when all seven gates pass, no admissible action remains, or the step budget
    is exhausted.}
  \label{fig:architecture}
\end{figure}

The benchmark \methodname{} loop comprises five modules:
SIMP solve, stress pass, evaluator, LLM/rule interpreter, and modifier
(Figure~\ref{fig:architecture}).  A separate Configurator can translate a
natural-language problem description into an initial specification in the full
workflow, but the quantitative benchmark in Section~\ref{sec:experiments} uses
predefined validated problem specifications for repeatability.  The
Configurator is therefore described for completeness but is not evaluated as a
benchmark variable.

\paragraph{Stage~1: One-time configuration.}
The \textbf{Configurator} (LLM-powered, Gemini 3.1 Flash-Lite Preview) translates a
natural-language problem description into a validated problem specification
$\mathcal{P}^{(0)}$ encoding domain geometry, boundary conditions, volume
fraction target, and optional passive or seed
regions~\cite{yang2026autosimp}.  This module runs once before the design
loop begins and produces the initial input to Stage~2.

\paragraph{Stage~2: Design iteration loop.}
Five modules execute in sequence at each design step
$t = 0, 1, \ldots, T_{\max}$:

\begin{enumerate}
  \item[(1)] \textbf{SIMP Solver} (numerical, no LLM).
    Receives the current specification $\mathcal{P}^{(t)}$ and runs
    compliance minimization using the SIMP method with an optimality-criteria (OC) update,
    three-field density filtering, and Heaviside ($\tanh$) projection.
The continuation schedule ramps penalization $p: 1 \to 4.5$, filter
    sharpness $\beta: 1 \to 32$, and tightens the minimum filter radius
    toward $r_{\min} = 1.2$ (Section~\ref{subsec:solver}).
    Outputs the converged density field
    $\bm{\rho}^{(t)} \in [0,1]^{n_e}$ and retained compliance
    $\Crep^{(t)}$.

  \item[(2)] \textbf{Stress Pass} (numerical, no LLM).
    Runs a forward FEA pass on the converged stiffness matrix and computes
    per-element von~Mises stress
    $\bm{\sigma}^{(t)} \in \mathbb{R}^{n_e}$
    (Section~\ref{subsec:stress}).  Both fields are rendered as
    $300 \times 300$~pixel PNG images
    $\mathcal{I}^{(t)} = \{\mathcal{I}^{(t)}_\rho,\,
    \mathcal{I}^{(t)}_\sigma\}$ for visual input to the Interpreter.

  \item[(3)] \textbf{Evaluator} (deterministic, no LLM).
    Applies seven gate checks and four additional diagnostics to the solver output
    (Section~\ref{subsec:evaluator}), producing an evaluation result
    $\mathcal{E}^{(t)}$ that records which gate checks passed or failed,
    including the added stress gate
    $\max_{e:\rho_e>0.5} \sigvm^{(e)} \leq \sigyield$.
    If all seven gate checks pass, the interpreter may stop with the current design as the
    final output $\bm{\rho}^{*}$.

  \item[(4)] \textbf{Interpreter} (hybrid LLM/rule module, Gemini 3.1 Flash-Lite Preview, $T{=}0$).
    Receives the density image $\mathcal{I}^{(t)}_\rho$, the stress
    image $\mathcal{I}^{(t)}_\sigma$, the evaluator check results
    $\mathcal{E}^{(t)}$, the current retained compliance $\Crep^{(t)}$, and the
    full action history $\{a^{(s)}\}_{s<t}$.  It computes deterministic
    stopping and solver-parameter safeguards, queries the LLM for
    vision-derived geometry suggestions, and returns a ranked list of
    candidate actions with structured JSON parameters from the vocabulary in
    Table~\ref{tab:actions}.
    Runs at temperature $T{=}0$ for reproducibility
    (Section~\ref{subsec:interpreter}).

  \item[(5)] \textbf{Modifier} (deterministic, no LLM).
    Applies the first valid action proposed by the Interpreter to produce
    the updated specification $\mathcal{P}^{(t+1)}$, which feeds the next
    SIMP Solver iteration.  Implements three seed kinds: soft
    (elevated-density warm start, the soft-seed condition), frozen solid
    (the passive-solid condition), or void lock (Section~\ref{subsec:seed}).
\end{enumerate}

The implementation terminates when the interpreter stop rule is satisfied,
when no admissible new action remains, or when the step budget
$T_{\max} = 5$ is exhausted.  The deterministic stop rule requires all seven
gate checks to pass, a compliance ratio below 1.05, and sufficient
load-path efficiency; repeated or blocked actions can also stop a trace
when further progress is not expected.  The final output is the
selected density field $\bm{\rho}^{*}$ and its retained compliance
$\Crep^{*}$.
In the full workflow, the optional Configurator and the
Interpreter invoke an LLM API. In the reported benchmark runs, the initial
specifications are predefined and only the Interpreter invokes the LLM API;
the SIMP Solver, Evaluator, and Modifier are purely numerical or rule-based.
The recorded traces store rendered inputs, parsed LLM action suggestions,
applied or rejected actions, and solver/evaluator states, but not complete
hosted-model input/output records or provider response identifiers; exact
hosted-model replay and independent reconstruction of the original model outputs are
therefore not available from the reproducibility materials.  The archived
records support code-level replay of the controller and solver state, but not
exact replay of those hosted-model calls.
Algorithm~\ref{alg:main} formalizes the iteration.

\begin{algorithm}[!htbp]
\caption{\methodname{} outer design loop}
\label{alg:main}
\begin{algorithmic}[1]
\Require Problem spec $\mathcal{P}^{(0)}$, step budget $T_{\max}$,
         yield threshold $\sigyield$
\Ensure  Selected density field $\bm{\rho}^{*}$, retained compliance $\Crep^{*}$
\State $\Crep^{*} \leftarrow \infty$;\; $t \leftarrow 0$
\While{$t < T_{\max}$}
  \State $\bm{\rho}^{(t)},\, \Crep^{(t)} \leftarrow \textsc{SimpSolve}(\mathcal{P}^{(t)})$
    \Comment{SIMP + Heaviside tail (Sec.~\ref{subsec:solver})}
  \State $\bm{\sigma}^{(t)} \leftarrow \textsc{StressPass}(\bm{\rho}^{(t)},\mathcal{P}^{(t)})$
    \Comment{one forward FEA (Sec.~\ref{subsec:stress})}
  \State $\mathcal{I}^{(t)} \leftarrow \textsc{RenderImages}(\bm{\rho}^{(t)},\bm{\sigma}^{(t)})$
    \Comment{$300{\times}300$\,px PNG pair}
  \State $\mathcal{E}^{(t)} \leftarrow \textsc{Evaluate}(\bm{\rho}^{(t)},\bm{\sigma}^{(t)},\sigyield)$
    \Comment{seven gate checks plus four diagnostics (Sec.~\ref{subsec:evaluator})}
  \If{\textsc{RetainedEligible}$(\bm{\rho}^{(t)},\Crep^{(t)})$ and $\Crep^{(t)} < \Crep^{*}$}
    \State $\Crep^{*} \leftarrow \Crep^{(t)}$;\; $\bm{\rho}^{*} \leftarrow \bm{\rho}^{(t)}$
  \EndIf
  \State $\mathcal{A}^{(t)},\,s^{(t)} \leftarrow
        \textsc{Interpret}(\mathcal{I}^{(t)}, \mathcal{E}^{(t)},
         \{a^{(s)}\}_{s<t})$
    \Comment{deterministic safeguards plus dual-image LLM geometry suggestions}
  \If{$s^{(t)} = \mathrm{stop}$ or $\mathcal{A}^{(t)} = \emptyset$}
    \State \textbf{break}
    \Comment{interpreter stop rule or no admissible new action}
  \EndIf
  \State $a^{(t)} \leftarrow \textsc{FirstAdmissible}(\mathcal{A}^{(t)})$
  \State $\mathcal{P}^{(t+1)} \leftarrow \textsc{Modify}(\mathcal{P}^{(t)}, a^{(t)})$
    \Comment{soft-seed injection (Sec.~\ref{subsec:seed})}
  \State $t \leftarrow t + 1$
\EndWhile
\State \textbf{return} $\bm{\rho}^{*},\, \Crep^{*}$
\end{algorithmic}

\vspace{3pt}
\footnotesize
\noindent\textsc{RetainedEligible} enforces the solver-side
finite-positive-compliance and grayness checks used by the retained-best
tracker.  The seven gate checks are reported separately unless a
feasibility-conditioned endpoint is explicitly used.
\end{algorithm}

\FloatBarrier

\subsection{SIMP solver and continuation schedule}
\label{subsec:solver}

The SIMP solver minimizes structural compliance
$C = \mathbf{u}^{\top}\mathbf{K}(\bm{\rho})\mathbf{u}$
subject to a volume constraint $V(\bm{\rho}) \leq v_f V_0$,
using the standard optimality-criteria (OC) density update with a three-field
(density--filtered density--projected density)
formulation~\cite{wang2011projection,lazarov2011filters}.
The element stiffness matrix is penalized as
$\mathbf{K}_e = \rho_e^p \mathbf{K}_e^0$, with $p$ ramped from 1.0 to 4.5
over the main solve.
Heaviside projection with sharpness parameter $\beta$ is doubled every
ten iterations after the penalty ramp, reaching $\beta = 32$ in the main
loop and held at $\beta = 32$ throughout the standardized 40-iteration
tail-refinement phase common to all conditions.
The filter radius $r_{\min}$ is tightened from its initial value toward
$r_{\min} = 1.2$ during the late main loop and fixed at that value for the tail.
Maximum solver iterations per design step: 120 (2D), 80 (3D).

Seed elements (Section~\ref{subsec:seed}) receive their initial density
assignment immediately before the first OC update of each solver call.  This
is a seed-biased reinitialization: each solver call starts from the current
volume-fraction density except that selected seed elements are initialized at
elevated density.  The subsequent OC dynamics are unconstrained.

\subsection{Stress field computation}
\label{subsec:stress}

After each SIMP solve, one additional forward FEA pass recovers the
full displacement field $\mathbf{U}$ under the same stiffness matrix.
Strains are evaluated at the element centroid using the $\mathbf{B}$-matrix
of the bilinear quadrilateral (2D, plane-stress) or trilinear hexahedral
(3D) element:
\begin{equation}
  \bm{\varepsilon}^{(e)} = \mathbf{B}_e \mathbf{u}_e,
\end{equation}
where $\mathbf{u}_e$ collects the nodal displacements of element $e$.
The constitutive modulus uses a fixed SIMP penalization exponent $p=3$
(independent of the solver's variable penalty schedule) to evaluate stress:
\begin{equation}
  E_e = E_{\min} + \rho_e^p \left( E_0 - E_{\min} \right),
  \quad E_{\min} = 10^{-9} E_0,
\end{equation}
so that void elements contribute negligible stress rather than creating
singular stress concentrations.
The 2D plane-stress von~Mises stress at element centroid is:
\begin{equation}
  \sigvm^{(e)} = \sqrt{
    \sigma_{xx}^2 + \sigma_{yy}^2 - \sigma_{xx}\sigma_{yy} + 3\tau_{xy}^2
  },
  \label{eq:vonmises}
\end{equation}
with $[\sigma_{xx},\,\sigma_{yy},\,\tau_{xy}]^\top = E_e\,\mathbf{D}\,\bm{\varepsilon}^{(e)}$
and $\mathbf{D}$ the plane-stress constitutive matrix.
For 3D elements we use
\begin{equation}
  \begin{aligned}
  \sigvm^{(e)}
  = \Bigg\{&
  \frac{1}{2}\left[
  (\sigma_{xx}-\sigma_{yy})^2+
  (\sigma_{yy}-\sigma_{zz})^2 \right.\\
  &\left.+
  (\sigma_{zz}-\sigma_{xx})^2\right]
  +3\left(\tau_{xy}^2+\tau_{yz}^2+\tau_{zx}^2\right)
  \Bigg\}^{1/2}.
  \end{aligned}
\end{equation}

The stress field is rendered as a blue-to-red (jet) colormap normalized
to the maximum $\sigvm^{(e)}$ over solid elements
($\rho_e > 0.5$). This matches the renderer used for the reported traces
and should be interpreted as a per-panel qualitative stress visualization
rather than an absolute cross-panel color scale.
All stress-field panels in this paper use this per-panel normalization unless
explicitly stated otherwise; color intensity should therefore not be compared
quantitatively across different panels.
Both the density image and the stress image are saved at
$300\times 300$~pixels (padded with white border) for the Interpreter query.
The yield threshold $\sigyield$ is calibrated from compliance-only
reference solves as described in Section~\ref{sec:experiments}.

\subsection{Dual-image LLM interpreter}
\label{subsec:interpreter}

At each design step $t$ the Interpreter submits to the LLM
(Gemini 3.1 Flash-Lite Preview, \texttt{gemini-3.1-flash-lite-preview},
temperature $T=0$) a structured prompt containing:
\begin{enumerate}[label=(\roman*)]
  \item the density field image $\mathcal{I}^{(t)}_\rho$,
  \item the von~Mises stress field image $\mathcal{I}^{(t)}_\sigma$,
  \item a JSON summary of the current evaluation: which gate checks passed
        or failed, the four recorded diagnostics, the current compliance
        $C^{(t)}$, retained compliance $\Crep^{*}$, and whether the
        maximum-stress gate is satisfied,
  \item the action history $\{a^{(s)}\}_{s<t}$ to prevent action repetition.
\end{enumerate}
The prompt instructs the LLM to: (a)~identify the primary structural
weakness visible in the density image (disconnected members, over-slender
sections, asymmetric load paths), (b)~locate the highest-stress region in
the stress image and estimate its domain coordinates in $[0,L_x]\times[0,L_y]$,
and (c)~output a ranked list of JSON action candidates with priority, action
type, and spatial parameters.
The implementation then validates those candidates, keeps LLM-derived geometry
actions ahead of duplicate deterministic geometry proposals, and appends
deterministic solver-parameter safeguards such as volume-fraction or
continuation-schedule changes when required by the evaluator state.  The same
deterministic rule layer supplies failed-check labels and the conservative stop
rule; the LLM is used for spatial interpretation rather than for overriding
validity criteria.

\paragraph{Action vocabulary.}
Table~\ref{tab:actions} lists the complete action vocabulary.
Actions retained from the earlier compliance-only LLM-SIMP controller
study~\cite{yang2026itersimp} operate on
global solver parameters; the three stress-aware actions operate on
the spatial density field and are the primary mechanism for hotspot targeting.

\begin{table}[!htbp]
  \caption{Complete \methodname{} action vocabulary. Actions marked
    $\dagger$ are introduced in this framework. Priority 1 = highest; the Interpreter
    proposes a ranked list and the Modifier applies the first valid action.}
  \label{tab:actions}
  \centering
  \footnotesize
  \setlength{\tabcolsep}{3.2pt}
  \begin{tabular}{llp{6.5cm}}
    \toprule
    Action & Type & Effect \\
    \midrule
    Change volume fraction          & global   & Increase or decrease target volume fraction \\
    Change filter radius            & global   & Tighten or loosen $r_{\min}$ \\
    Change penalization             & global   & Adjust SIMP penalty exponent $p$ \\
    Insert passive void             & spatial  & Insert a passive void (locks $\rho_e = \rho_{\min}$) \\
    Insert frozen solid             & spatial  & Insert a frozen solid ($\rho_e = 1$, ablation only) \\
    \midrule
    Reinforce stress hotspot$^\dagger$ & spatial & Place a soft-seed region centered on the peak-stress solid element; in reported 3D traces, circular seed regions are extruded through depth (default $\rhobar_{\mathrm{seed}}=0.85$; reported LLM traces use 0.85--0.95) \\
    Widen stressed member$^\dagger$    & spatial & Add an overlapping chain of soft-seed circles along a stressed member's centerline (default $\rhobar_{\mathrm{seed}}=0.80$; reported LLM traces use 0.80--0.90) \\
    Redistribute material$^\dagger$    & spatial & Void a low-stress region; simultaneously soft-seed the corresponding high-stress region (default $\rhobar_{\mathrm{seed}}=0.85$; reported LLM traces use 0.85--0.95) \\
    \bottomrule
  \end{tabular}
\end{table}

\paragraph{Stress-gate fallback.}
When the stress gate check fails ($\max_e\sigvm^{(e)} > \sigyield$) and
no spatial action can plausibly resolve it within the remaining step budget,
the Interpreter may propose a volume-fraction increase to reduce global stress
levels before resuming spatial corrections. The Modifier can also compound a
spatial action with compatible global changes such as volume-fraction or
filter-radius adjustments. Consequently, the reported results are
controller-policy outcomes rather than isolated tests of a single spatial seed
action.

\subsection{Soft density seed mechanism}
\label{subsec:seed}

The spatial actions in Table~\ref{tab:actions} produce circular or
rectangular geometry regions in the problem specification $\mathcal{P}$.
Each region carries a \emph{kind} attribute that determines how the SIMP
solver treats the enclosed elements.  Three kinds are defined:

\begin{enumerate}
  \item[(1)] \textbf{Void} ($\rho_e = \rho_{\min}$, frozen):
        elements are locked to near-zero density throughout the solve and
        excluded from the OC update.  Used for holes, cutouts, and
        load-clearance zones.

  \item[(2)] \textbf{Solid} ($\rho_e = 1$, frozen):
        elements are locked to full density and excluded from the OC
        update.  Material is forced into the region regardless of
        sensitivity.  Used only in the passive-solid condition to quantify
the cost of removing optimizer freedom.  These frozen elements are
        counted in the reported physical volume-fraction check, but they
        are not redistributed by the OC bisection; passive-solid results
therefore change the optimizer's feasible design space and are
        treated as an ablation rather than the main soft-seed mechanism.

  \item[(3)] \textbf{Seed} ($\rho_e \leftarrow \rhobar_{\mathrm{seed}}$,
        \emph{not frozen --- our method}):
        elements are initialized at elevated density immediately before
        the first OC update of the next solver call, then participate in
        the standard OC update identically to all other elements.
        The seed density is \emph{not} enforced at any subsequent
iteration; the optimizer is free to redistribute material away
        from the seeded region if sensitivity analysis determines it
        carries insufficient load.
\end{enumerate}

Formally, let $\mathcal{S}^{(t)} \subseteq \{1,\ldots,n_e\}$ be the set
of seed elements proposed at step $t$.
The modified OC initialization is:
\begin{equation}
  \rho_e^{(0)} \leftarrow
  \begin{cases}
    \rhobar_{\mathrm{seed}} & e \in \mathcal{S}^{(t)}, \\
  v_f                     & \text{otherwise (uniform re-initialization)},
  \end{cases}
  \label{eq:seed_init}
\end{equation}
and for all $k \geq 1$ the standard OC update applies uniformly:
\begin{equation}
  \rho_e^{(k+1)} = \min\!\left(1,\max\!\left(\rho_{\min},\,
    \rho_e^{(k)} \cdot \left(\frac{-\partial C/\partial\rho_e}{\lambda}\right)^{\!\eta}
  \right)\right),
  \label{eq:oc_update}
\end{equation}
where $\lambda$ is the Lagrange multiplier for the volume constraint found
by bisection, and $\eta = 0.5$ is the OC damping exponent.
In implementation, this OC candidate is clipped by the move limit $m$:
$\rho_e^{(k+1)} \in [\rho_e^{(k)} - m,\, \rho_e^{(k)} + m]$, with
$m=0.05$ in the standardized tail-refinement phase.
Crucially, $e \in \mathcal{S}^{(t)}$ receives no special treatment in
Equation~\eqref{eq:oc_update}: the seed-biased density assignment evolves
freely under sensitivity analysis from iteration $k=1$ onward.

This differs from the passive-solid ablation, in
which seeded elements are excluded from Equation~\eqref{eq:oc_update}
entirely and contribute to the global stiffness at full modulus $E_0$
regardless of sensitivity.
The practical consequence is that a mis-placed solid seed imposes a
geometric constraint that the optimizer cannot escape, whereas a soft seed
merely biases the initialization: if the sensitivity analysis determines
that the seeded region carries insufficient load, material will be
redistributed away from it within a few OC iterations.

\begin{figure}[htbp]
  \centering
  \includegraphics[width=\linewidth]{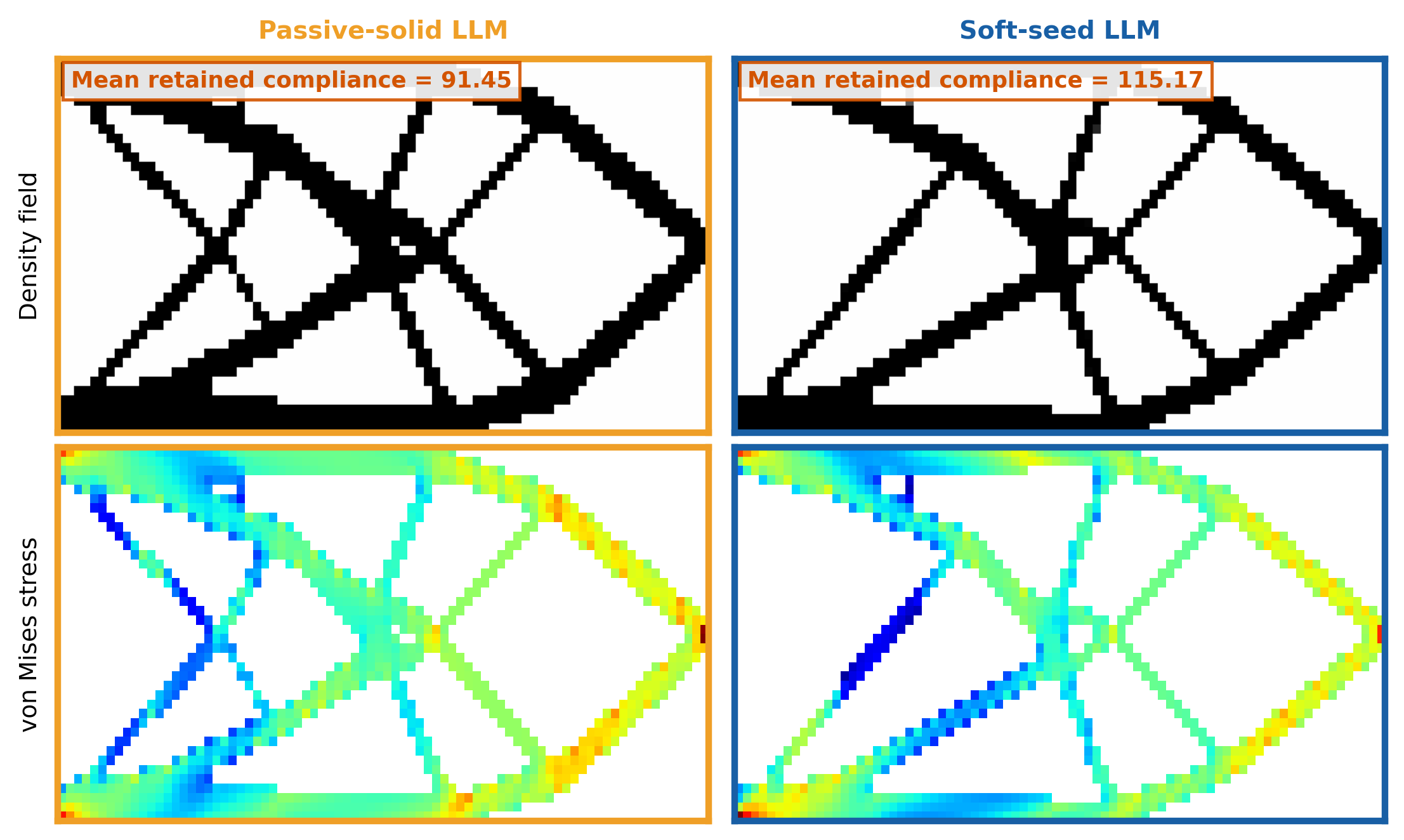}
  \caption{Passive-solid seeding (left) vs.\ soft density seeding
    (right) for the cantilever-with-two-voids benchmark.
    Top row: retained-best density field. Bottom row: von~Mises stress field.
    The passive-solid topology (amber border, $C=91.5$) has lower
    compliance than the soft-seed topology (blue border, $C=115.2$) in the
    reported experiment.
    The two panels are seed-42 examples from the same benchmark and compare
    the retained passive-solid and soft-seed outcomes. They should be read as
    an illustrative mechanism comparison rather than a controlled same-location
    intervention, because the realized action sequences differ between the two
    traces.}
  \label{fig:seed_mechanism}
\end{figure}

Figure~\ref{fig:seed_mechanism} illustrates the contrast on
the cantilever-with-two-voids benchmark.
In this case the passive-solid intervention (left, $C=91.5$) outperforms
the soft-seed intervention (right, $C=115.2$), even though the latter
preserves optimizer freedom.  The example therefore illustrates both the
mechanical distinction between the two materialization modes and the
reason the passive-solid variant must be reported explicitly rather than
treated as a secondary implementation detail.

\paragraph{Choice of $\rhobar_{\mathrm{seed}}$.}
The seed density must be high enough to bias the initial sensitivity
$\partial C/\partial\rho_e$ toward retaining material in the hotspot region
(requiring $\rhobar_{\mathrm{seed}} > v_f$ for any problem with $v_f < 0.85$),
yet not so high as to approach $\rho_e = 1$ and effectively lock the element
before the first OC step. We use 0.85 as the default hotspot and
redistribution value because it is above every initial benchmark volume
fraction while remaining below the frozen-solid limit. The widen-member
action uses a slightly lower default seed density
($\rhobar_{\mathrm{seed}} = 0.80$) to avoid excessive material commitment
along extended flanking regions. In the reported LLM traces, the
Interpreter could propose the seed-density parameter; realized values range
from 0.80 to 0.95. This parameter variation is part of the reported
controller policy, not an isolated fixed-seed-density ablation.

\subsection{Stress-gated evaluator}
\label{subsec:evaluator}

After each solver call the Evaluator applies seven \emph{gate checks}
(a failed gate check marks the run as not yet passing and drives the next
Interpreter query) and records four \emph{informational diagnostics}
(reported but not used for gating).
Table~\ref{tab:checks} lists the gate checks and diagnostics.

\begin{table}[!htbp]
  \caption{Evaluator check suite. Gate checks must all pass for a run to be
    declared converged. Informational checks are reported but do not gate
    the loop. Checks marked $\dagger$ are introduced in this framework.}
  \label{tab:checks}
  \centering
  \begingroup
  \footnotesize
  \setlength{\tabcolsep}{3pt}
  \begin{tabularx}{\linewidth}{@{}>{\raggedright\arraybackslash}p{0.23\linewidth}l>{\raggedright\arraybackslash}p{0.16\linewidth}>{\raggedright\arraybackslash}X@{}}
    \toprule
    Check & Gate? & Threshold & Definition \\
    \midrule
    Load outside void     & Gate & \textemdash{}       & No point load inside a passive void region \\
    Connectivity            & Gate & $\geq 0.99$ & Fraction of solid elements in largest 8-connected component \\
    Compliance ratio       & Gate & $\leq 2.0$  & $\Cfinal / \Crep$ \\
    Grayness                & Gate & $\leq 0.15$ & $\mathrm{mean}_e\,\min(\rho_e,1-\rho_e) / 0.25$ \\
    Volume fraction        & Gate & $|\Delta v_f| \leq 0.05$ & Absolute deviation from target $v_f$ \\
    Convergence             & Gate & $n_{\mathrm{iter}} \leq 1.5\,n_{\max}$ & Total solver iterations \\
    \textbf{Maximum stress}$^\dagger$     & Gate & $\leq \sigyield$ & $\max_{e:\rho_e>0.5} \sigvm^{(e)}$ \\
    \midrule
    Thin members           & Info & \textemdash{}       & Fraction of solid boundary elements (erosion proxy) \\
    Checkerboard pattern            & Info & \textemdash{}       & Fraction of solid elements differing from all 4-neighbours \\
    Load-path efficiency  & Info & \textemdash{}       & Solid fraction heuristic on direct load path \\
    \textbf{Stress concentration}$^\dagger$ & Info & \textemdash{} & $\mathrm{SCF} = \max_e\sigvm^{(e)} / \mathrm{mean}_e\sigvm^{(e)}$ over solid elements \\
    \bottomrule
  \end{tabularx}
  \endgroup
\end{table}

The \emph{maximum-stress} gate implements the primary stress constraint:
\begin{equation}
  \max_{e:\,\rho_e > 0.5} \sigvm^{(e)} \;\leq\; \sigyield.
  \label{eq:stress_gate}
\end{equation}
Only solid elements ($\rho_e > 0.5$) enter this gate, excluding near-void
elements where the SIMP-penalized modulus $E_e \approx E_{\min}$ would
produce unreliably small stresses.
The \emph{stress-concentration factor} (SCF) is reported as an
informational diagnostic: a high SCF signals a localized hotspot that
the Interpreter should address at the next step, even when the gate passes.

\section{Experimental Setup}
\label{sec:experiments}

\subsection{Benchmark problems}

Table~\ref{tab:problems} lists all 22 benchmark problems.

\begin{table}[!htbp]
  \caption{Benchmark problem suite. All 2D problems use plane-stress
    elements; 3D problems use trilinear hexahedra. vf~= volume fraction.
    Benchmark label mappings are listed in Appendix~\ref{app:repro_map}.}
  \label{tab:problems}
  \centering
  \begingroup
  \footnotesize
  \setlength{\tabcolsep}{3pt}
  \begin{tabularx}{\linewidth}{@{}r>{\raggedright\arraybackslash}p{0.24\linewidth}>{\centering\arraybackslash}p{0.13\linewidth}r>{\centering\arraybackslash}p{0.045\linewidth}>{\raggedright\arraybackslash}X@{}}
    \toprule
    ID & Name & Mesh & $n_e$ & vf & Boundary conditions \\
    \midrule
    \multicolumn{6}{l}{\textit{2D benchmark (16 problems)}} \\
    1  & Sparse bridge          & $80{\times}20$  & 1600 & 0.18 & Pin--roller, distributed top load \\
    2  & Bridge with circular void & $80{\times}20$  & 1600 & 0.25 & Pin--roller, top load, circular void \\
    3  & Cantilever with two voids & $80{\times}40$  & 3200 & 0.25 & Left fixed, mid-right load, 2 voids \\
    4  & Low-volume cantilever  & $60{\times}30$  & 1800 & 0.12 & Left fixed, mid-right load \\
    5  & Asymmetric low-volume cantilever & $90{\times}30$  & 2700 & 0.18 & Left fixed, off-axis load \\
    6  & Central-void frame     & $60{\times}60$  & 3600 & 0.22 & Bottom fixed, top-center load, central void \\
    7  & L-bracket              & $40{\times}40$  & 1600 & 0.30 & Top fixed, bottom-right load, upper-right void \\
    8  & T-bracket              & $40{\times}60$  & 2400 & 0.25 & Bottom fixed, right load, left void \\
    9  & Simply supported beam  & $80{\times}20$  & 1600 & 0.30 & Two pin-y supports, mid-top load \\
    10 & Portal frame           & $60{\times}40$  & 2400 & 0.22 & Two corner fixed, distributed top load, central void \\
    11 & Asymmetric MBB beam    & $90{\times}30$  & 2700 & 0.30 & Left pin-x, right bottom pin-y, off-center load \\
    12 & Deep beam              & $60{\times}30$  & 1800 & 0.35 & Pin--roller, distributed top load \\
    13 & Michell truss          & $80{\times}40$  & 3200 & 0.15 & Pin--roller, mid-top point load \\
    14 & Basic cantilever       & $60{\times}30$  & 1800 & 0.35 & Left fixed, mid-right load \\
    15 & MBB beam               & $90{\times}30$  & 2700 & 0.35 & Left pin-x, right bottom pin-y, top-left load \\
    16 & Dual-load cantilever   & $60{\times}30$  & 1800 & 0.30 & Left fixed, two right-edge loads \\
    \midrule
    \multicolumn{6}{l}{\textit{3D benchmark (6 problems)}} \\
    17 & 3D cantilever          & $40{\times}20{\times}10$ & 8000 & 0.25 & Left face fixed, mid-right load \\
    18 & 3D MBB beam            & $40{\times}14{\times}6$  & 3360 & 0.18 & Left pin-x, right bottom pin-y \\
    19 & 3D bridge              & $30{\times}10{\times}6$  & 1800 & 0.18 & Pin--roller, distributed top load \\
    20 & 3D L-bracket           & $20{\times}20{\times}5$  & 2000 & 0.25 & Top fixed, bottom-right load, void \\
    21 & 3D Michell truss       & $24{\times}12{\times}6$  & 1728 & 0.12 & Pin--roller, mid-top load \\
    22 & 3D torsion block       & $20{\times}10{\times}8$  & 1600 & 0.18 & Left fixed, opposing off-axis loads \\
    \bottomrule
  \end{tabularx}
  \endgroup
\end{table}

\subsection{Conditions}

The controller-policy diagnostic compares three conditions:

\begin{enumerate}
  \item[(1)] \textbf{Soft-seed LLM (IterSIMP-$\sigma$):} LLM interpreter with
    dual-image input and soft density seeding, starting from predefined
    benchmark specifications.
  \item[(2)] \textbf{Rule-based:} identical SIMP solver and tail schedule;
    no LLM; actions are heuristic scalar adjustments
(change volume fraction, change filter radius, change penalization).
  \item[(3)] \textbf{Passive-solid LLM:} LLM interpreter with dual-image input,
    but seed regions are frozen as passive solid rather than soft-seeded.
\end{enumerate}

The fixed-volume 2D attribution study adds six further comparisons.  These
are exact numerical hotspot seeding, random regional seeding, density-only
input, stress-only input, numeric-only input, and global LLM action control.
The exact-hotspot condition uses the numerical von~Mises stress array rather
than rendered images to place the same soft-seed intervention around the
maximum-stress element.
The random condition samples stress-region seed locations with the same
spatial-action budget.  The input ablations keep the LLM controller but remove
parts of the multimodal context or spatial-action freedom.  These fixed-volume
results are reported separately from the three-condition controller-policy comparison
because they address problem-change and localization confounds directly and
use a separate 2D stress-threshold calibration.

All conditions use identical SIMP solver settings: maximum 120 solver
iterations per design step (capped at 80 for 3D), up to 5 design steps
per problem, and a standardized tail-refinement schedule (penalty
$p=4.5$, $r_{\min}=1.2$, move~$=0.05$, $\beta=32$, 40 tail iterations).
The volume fraction in Table~\ref{tab:problems} is the initial benchmark
    target.  Because changing the volume-fraction target is part of the
controller action vocabulary, the reported compliance values are outcomes
of the corresponding controller policies, not fixed-volume ablations,
unless explicitly labeled otherwise.  Final volume targets are reported in
the 3D table and mapped for all 2D primary traces in
Appendix~\ref{app:repro_map}.
Because the benchmark suite is synthetic and author-constructed, it should be
read as a controlled stress-gate test bed rather than as a representative
sample of all structural design settings.

\subsection{Stress threshold calibration}
\label{subsec:calibration}

The yield threshold $\sigyield$ is calibrated once before the main
experiment.  Each of the 16 two-dimensional problems is solved in
compliance-only mode (no stress constraint, single run), and the peak
von~Mises stress $\max_e \sigvm^{(e)}$ is recorded for each problem.
The threshold is then set as a percentile of this distribution:
$\sigyield = \mathrm{pct}_q\!\bigl(\max_e \sigvm^{(e)}\bigr)$
over the 16 problems.  At $q = 50$ (the primary evaluation point)
this yields $\sigyield = 120.3$ in normalized stress units (consistent
with $E_0 = 1$; all quantities are nondimensionalized).
This calibration ensures that approximately half the problems fail the
stress gate on a compliance-only solve, creating a benchmark where
stress-aware intervention has room to contribute without being either
trivially satisfied or impossibly tight.

For the 3D benchmark reported here, we retain the same paper-wide
threshold $\sigyield = 120.3$.  This keeps the scalar tables under a
single stress gate, but it also means that the 3D results should be
interpreted as exploratory controller evidence rather than a separately
calibrated stress-active 3D study.

The sensitivity of results to the choice of $\sigyield$ is characterized
across six percentile levels ($q = 40, 45, 50, 55, 60, 70$) in
Section~\ref{sec:results}.

\subsection{Statistical protocol}

The primary evaluation is conducted at $\sigyield = 120.3$ (50th
percentile).
Each of the two primary conditions (soft-seed LLM and rule-based) is repeated
with three recorded run seeds (42, 123, 7).  The current solver initializes
most runs from a uniform volume-fraction density; the seeds are retained for
run identity and for solver/controller paths that use seeded stochastic state.
With deterministic LLM decoding, identical action traces can occur across
repeat seeds, so the repeat count should be read as trace-level replication
rather than independent LLM sampling.
Statistical significance is reported using the two-sided Wilcoxon signed-rank
test as the primary inferential check.  The corresponding one-sided value in
the expected improvement direction is reported only as a directional
diagnostic.
Problems with coefficient of variation $>10\%$ are disclosed individually
(Section~\ref{subsec:variance}).
Sensitivity to $\sigyield$ is characterized across six percentile levels
using single runs per condition.
Per-problem comparisons are descriptive and are not adjusted for multiple
testing.

\paragraph{Retained compliance metric.}
The scalar result tables report $\Crep$, the compliance value used for
comparisons after retaining the best compliance observed across design steps.
It is not necessarily the final solver state and, in several 2D cases, it is
not a final state satisfying all seven gate checks.  We therefore report final
gate-pass counts separately from the compliance metric.  The final-state compliance is
retained separately as $\Cfinal$.  When a final logged state satisfies the same
finite-positive, grayness, and gate checks as the retained tracker, it is
eligible for the reported retained-best value; the 3D bridge rule row below is
corrected on that basis.
Unless explicitly stated otherwise, scalar tables, classification counts, geometric means,
and figure labels use $\Crep$.  Standard deviations are descriptive population
standard deviations computed over the three repeat-level retained-best
compliance values, not inferential uncertainty estimates.  A retained candidate
must also satisfy the solver-side finite-positive-compliance and grayness
eligibility checks used by the retained-best tracker.
Because the recorded traces were not generated under a
feasibility-conditioned selection rule, $\Crep$ is a diagnostic
controller-policy metric rather than a feasible-final primary performance
metric.  The fixed-volume study in Section~\ref{subsec:fixed_volume_study}
therefore reports feasibility-conditioned retained compliance and
feasible-final compliance as primary endpoints.  Appendix~\ref{app:repro_map}
maps these reader-facing metrics to the archived summary fields used for
reproduction.

\section{Results}
\label{sec:results}

\subsection{2D controller-policy benchmark at \texorpdfstring{$\sigyield = 120.3$}{sigma = 120.3}}

In 2D, soft-seed control yields a small non-significant aggregate reduction in
retained compliance, with strong problem dependence.  Because the comparison is
neither fixed-volume nor feasibility-conditioned, this subsection is a
controller-policy diagnostic rather than a same-volume performance comparison.
Table~\ref{tab:main_results} reports per-problem retained-best compliance
(mean~$\pm$~population std,
$N=3$ repeats) for all 16 two-dimensional problems at the primary
evaluation point $\sigyield = 120.3$ (50th percentile), together with the
number of repeats whose final evaluator state passes all seven gate checks. These
values are controller-policy outcomes; several traces change the final
volume-fraction target, as summarized in Appendix~\ref{app:repro_map}.

\begin{table}[!htbp]
  \caption{Per-problem retained-best compliance $\Crep$ at
    $\sigyield=120.3$ (calibration percentile $q=50$),
    $N=3$ repeats per condition (seeds 42, 123, 7).
    Std columns use population standard deviation.
    Pass columns give the number of repeats whose final state passes all
    seven gate checks; four additional diagnostics are reported but do not
    determine validity.
    $\Delta$: relative retained-compliance difference
    $= (C_{\text{rule}} - C_{\text{LLM}})/C_{\text{rule}} \times 100\%$.}
  \label{tab:main_results}
  \centering
  \scriptsize
  \setlength{\tabcolsep}{2.8pt}
  \begin{tabular}{p{3.05cm}rrrrrrrl}
    \toprule
    Problem & LLM mean & pop. std & LLM pass & Rule mean & pop. std & Rule pass & $\Delta$ (\%) & Guarded class \\
    \midrule
    Asymmetric low-volume cantilever & 173.63 & 11.04 & 0/3 & 185.77 & 0.00 & 0/3 & $+6.5$  & LLM \\
    Sparse bridge         & 371.89 & 89.00 & 0/3 & 293.40 & 0.00 & 0/3 & $-26.8$ & Tie \\
    Bridge with circular void & 230.87 & 12.29 & 0/3 & 250.82 & 0.00 & 0/3 & $+8.0$  & LLM \\
    Low-volume cantilever &  99.68 &  4.50 & 3/3 & 138.48 & 0.00 & 3/3 & $+28.0$ & LLM \\
    Basic cantilever      &  94.76 &  0.00 & 3/3 &  94.76 & 0.00 & 3/3 & $+0.0$  & Tie \\
    Cantilever with two voids & 115.17 &  1.01 & 3/3 & 117.19 & 0.00 & 3/3 & $+1.7$  & Tie \\
    Deep beam             &  35.08 &  0.00 & 3/3 &  35.08 & 0.00 & 3/3 & $+0.0$  & Tie \\
    Dual-load cantilever  & 288.18 &  0.74 & 3/3 & 291.13 & 0.00 & 3/3 & $+1.0$  & Tie \\
    Central-void frame    &  17.61 &  0.00 & 3/3 &  18.26 & 0.00 & 3/3 & $+3.5$  & LLM \\
    L-bracket             &  93.92 &  0.00 & 3/3 &  93.99 & 0.00 & 3/3 & $+0.1$  & Tie \\
    Asymmetric MBB beam   & 117.96 &  5.20 & 0/3 & 114.30 & 0.00 & 0/3 & $-3.2$  & Tie \\
    MBB beam              & 154.49 &  0.02 & 0/3 & 154.18 & 0.00 & 0/3 & $-0.2$  & Tie \\
    Michell truss         &  26.05 &  3.16 & 2/3 &  29.64 & 0.00 & 0/3 & $+12.1$ & LLM \\
    Portal frame          &  89.81 & 10.80 & 2/3 &  81.54 & 0.00 & 3/3 & $-10.1$ & Tie \\
    Simply supported beam &  66.26 &  0.00 & 3/3 &  66.26 & 0.00 & 3/3 & $+0.0$  & Tie \\
    T-bracket             &  11.88 &  0.10 & 3/3 &  11.95 & 0.00 & 3/3 & $+0.6$  & Tie \\
    \midrule
    \multicolumn{9}{l}{\textbf{Raw lower means: LLM 9; rule-based 4; equal 3}} \\
    \multicolumn{9}{l}{\textbf{Guarded class after 3\%/std guard: lower LLM 5; lower rule-based 0; ties 11}} \\
    \multicolumn{6}{l}{Geo.\ mean $C_{\text{LLM}}/C_{\text{rule}}$} & $= 0.981$ & & \\
    \bottomrule
  \end{tabular}
  \medskip

  \noindent
  \small Retained classes are descriptive; raw means and final gate-pass
  counts should be interpreted separately.  The classes use the practical-tolerance and standard-deviation
  guard defined in the surrounding text; several mean differences, including three cases with
  lower rule-based means, are therefore reported as guarded ties.  This
  classification is descriptive and does not override the raw mean
  differences.  Pass counts are not used to assign classes; they separate
  retained-compliance comparisons from final-state gate-check status.  Retained-best
  states may fail final gate checks, and the controller policies may
  change the final volume-fraction target.
\end{table}

Several retained-compliance differences are therefore coupled to large
controller-induced volume-fraction changes; Appendix~\ref{app:repro_map}
reports final volume targets, with some cases changing by more than 0.2
absolute volume fraction.  This material-budget coupling is why the
fixed-volume attribution study is reported separately.

\paragraph{Feasibility-conditioned re-score of existing traces.}
Table~\ref{tab:feasible_existing} re-scores the same recorded traces under
stricter feasibility-conditioned metrics.  Runs that
never pass all seven gate checks are marked infeasible and are excluded from
the corresponding paired mean rather than contributing a low retained
compliance.  This re-score checks existing traces; it is not a rerun under a
feasibility-conditioned selection rule.  It reduces the comparable sample: for
$C_{\mathrm{feas}}$, five of the 16 problem pairs are incomplete; for
$C_{\mathrm{final,feas}}$, six pairs are incomplete.  On the comparable subset,
the paired mean counts are lower LLM 7, lower rule 1, equal 3 for feasible
retained compliance, and lower LLM 6, lower rule 2, equal 2 for feasible-final
compliance.

\begin{table*}[t]
  \caption{Existing-trace feasibility-conditioned compliance audit for the 16 two-dimensional primary traces. $C_{\mathrm{feas}}$ is the lowest retained compliance among recorded steps passing all seven gate checks. $C_{\mathrm{final,feas}}$ is the final-state compliance when the final state passes all seven gate checks. Entries marked ``--'' are infeasible under that scoring rule and do not contribute to paired means. The repeats column reports feasible-any/final-feasible counts out of three.}
  \label{tab:feasible_existing}
  \centering
  \scriptsize
  \setlength{\tabcolsep}{2.7pt}
  \begin{tabular}{p{3.15cm}rrcrrc}
    \toprule
    Problem & LLM $C_{\mathrm{feas}}$ & LLM $C_{\mathrm{final,feas}}$ & LLM reps & Rule $C_{\mathrm{feas}}$ & Rule $C_{\mathrm{final,feas}}$ & Rule reps \\
    \midrule
    Asymmetric low-volume cantilever & -- & -- & 0/0 & -- & -- & 0/0 \\
    Sparse bridge & -- & -- & 0/0 & -- & -- & 0/0 \\
    Bridge with circular void & -- & -- & 0/0 & -- & -- & 0/0 \\
    Low-volume cantilever & 99.68 & 99.83 & 3/3 & 138.48 & 139.07 & 3/3 \\
    Basic cantilever & 94.76 & 94.76 & 3/3 & 94.76 & 94.76 & 3/3 \\
    Cantilever with two voids & 115.17 & 115.21 & 3/3 & 117.19 & 117.54 & 3/3 \\
    Deep beam & 35.08 & 33.19 & 3/3 & 35.08 & 33.12 & 3/3 \\
    Dual-load cantilever & 288.18 & 288.25 & 3/3 & 291.13 & 291.13 & 3/3 \\
    Central-void frame & 17.61 & 21.93 & 3/3 & 18.26 & 33.44 & 3/3 \\
    L-bracket & 93.92 & 93.92 & 3/3 & 93.99 & 93.99 & 3/3 \\
    Asymmetric MBB beam & -- & -- & 0/0 & -- & -- & 0/0 \\
    MBB beam & -- & -- & 0/0 & -- & -- & 0/0 \\
    Michell truss & 26.05 & 26.55 & 3/2 & 29.64 & -- & 3/0 \\
    Portal frame & 89.81 & 98.54 & 3/2 & 81.54 & 81.54 & 3/3 \\
    Simply supported beam & 66.26 & 66.92 & 3/3 & 66.26 & 66.92 & 3/3 \\
    T-bracket & 11.88 & 15.04 & 3/3 & 11.95 & 16.16 & 3/3 \\
    \midrule
    \multicolumn{7}{l}{Retained-any raw lower means: LLM 9; rule-based 4; equal 3.} \\
    \multicolumn{7}{l}{Feasible-retained paired means, excluding infeasible pairs: LLM lower 7; rule lower 1; equal 3; incomplete 5.} \\
    \multicolumn{7}{l}{Final-feasible paired means, excluding infeasible pairs: LLM lower 6; rule lower 2; equal 2; incomplete 6.} \\
    \bottomrule
  \end{tabular}
\end{table*}

\paragraph{Statistical test.}
For the soft-seed LLM condition, a Wilcoxon signed-rank test on the 16
per-problem mean compliance pairs gives $W=33$, two-sided $p=0.382$
(one-sided $p=0.191$).  The geometric mean is below one, but this paired
test does not provide statistical evidence of a soft-seed advantage.

\paragraph{Guarded descriptive classification.}
A problem is classified as \emph{lower LLM retained compliance} when the LLM mean compliance
is strictly lower than the rule-based mean and the absolute difference
$|C_{\text{rule}} - C_{\text{LLM}}|$ exceeds the larger of the two
conditions' standard deviations, i.e., $\max(\text{std}_{\text{LLM}},
\text{std}_{\text{rule}})$, and also exceeds a practical 3\% compliance
tolerance.
Problems where the mean difference does not exceed this threshold are
classified as ties.
Under this criterion, the low-percentage differences for the L-bracket,
cantilever with two voids, dual-load cantilever, MBB beam, and T-bracket are
treated as practical ties.  Three problems have lower rule-based mean
compliance (sparse bridge, portal frame, and asymmetric MBB beam), but these
are also classified as ties because the differences do not exceed the
soft-seed repeat-to-repeat spread.

\begin{figure}[htbp]
  \centering
  \includegraphics[width=\linewidth]{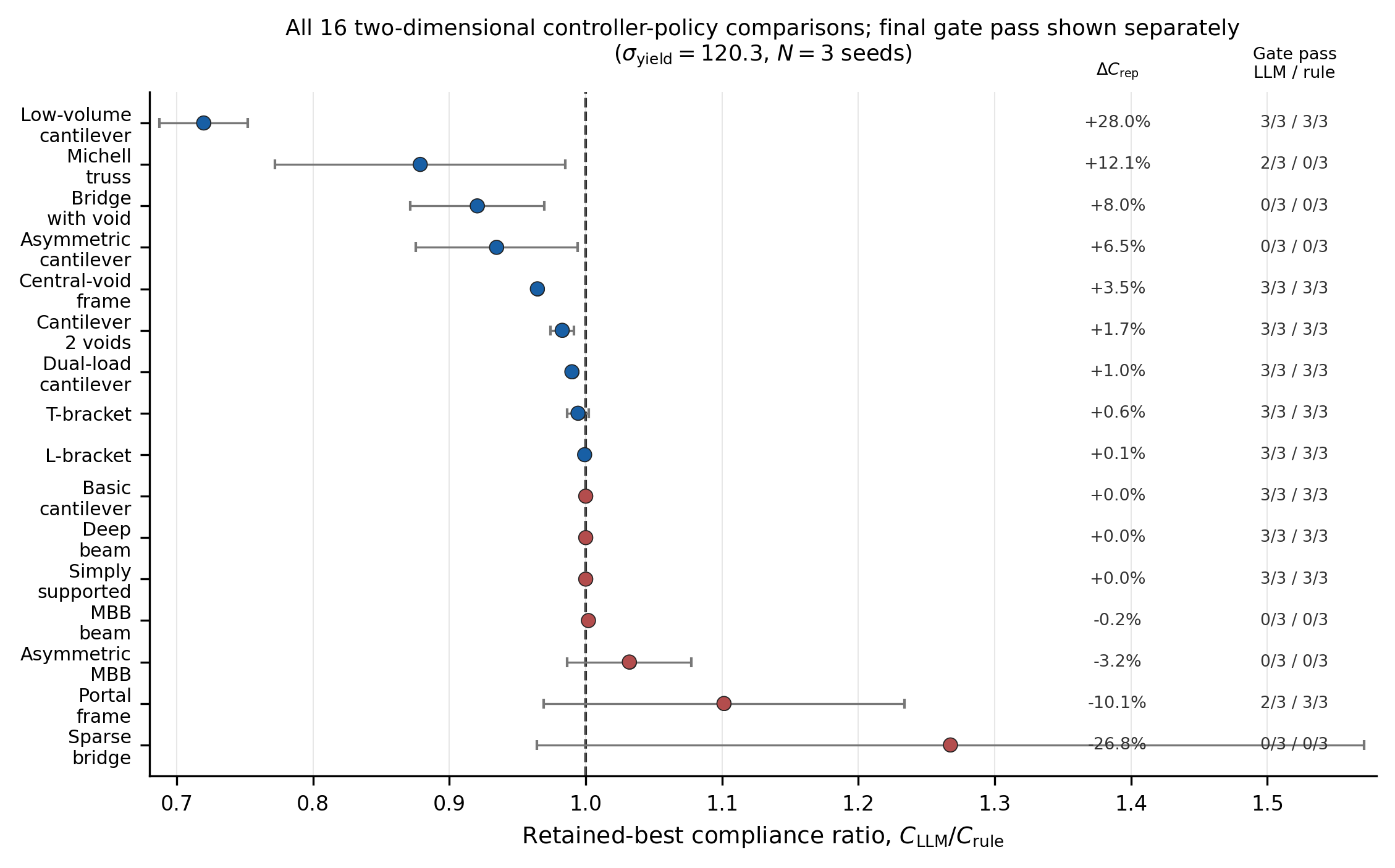}
  \caption{Retained-best compliance ratio for all 16 two-dimensional
    controller-policy comparisons at $\sigyield = 120.3$.
    Points show $C_{\mathrm{LLM}}/C_{\mathrm{rule}}$ from three-seed mean
    retained-best compliance; horizontal intervals show the descriptive
    soft-seed population standard deviation scaled by the rule mean.  Values
    left of one have lower LLM retained compliance, values right of one have
    lower rule-based retained compliance.  Right-side columns report relative
    retained-compliance difference and final gate-pass counts for LLM/rule
    repeats.}
  \label{fig:compliance_bars}
\end{figure}

Figure~\ref{fig:compliance_bars} visualizes all 16 retained-compliance ratios,
sorted by effect size, with final gate-pass counts annotated because not all
retained-best states are final gate-passing states and some policies change
final volume fraction.

\begin{figure}[p]
  \centering
  \includegraphics[width=0.92\linewidth]{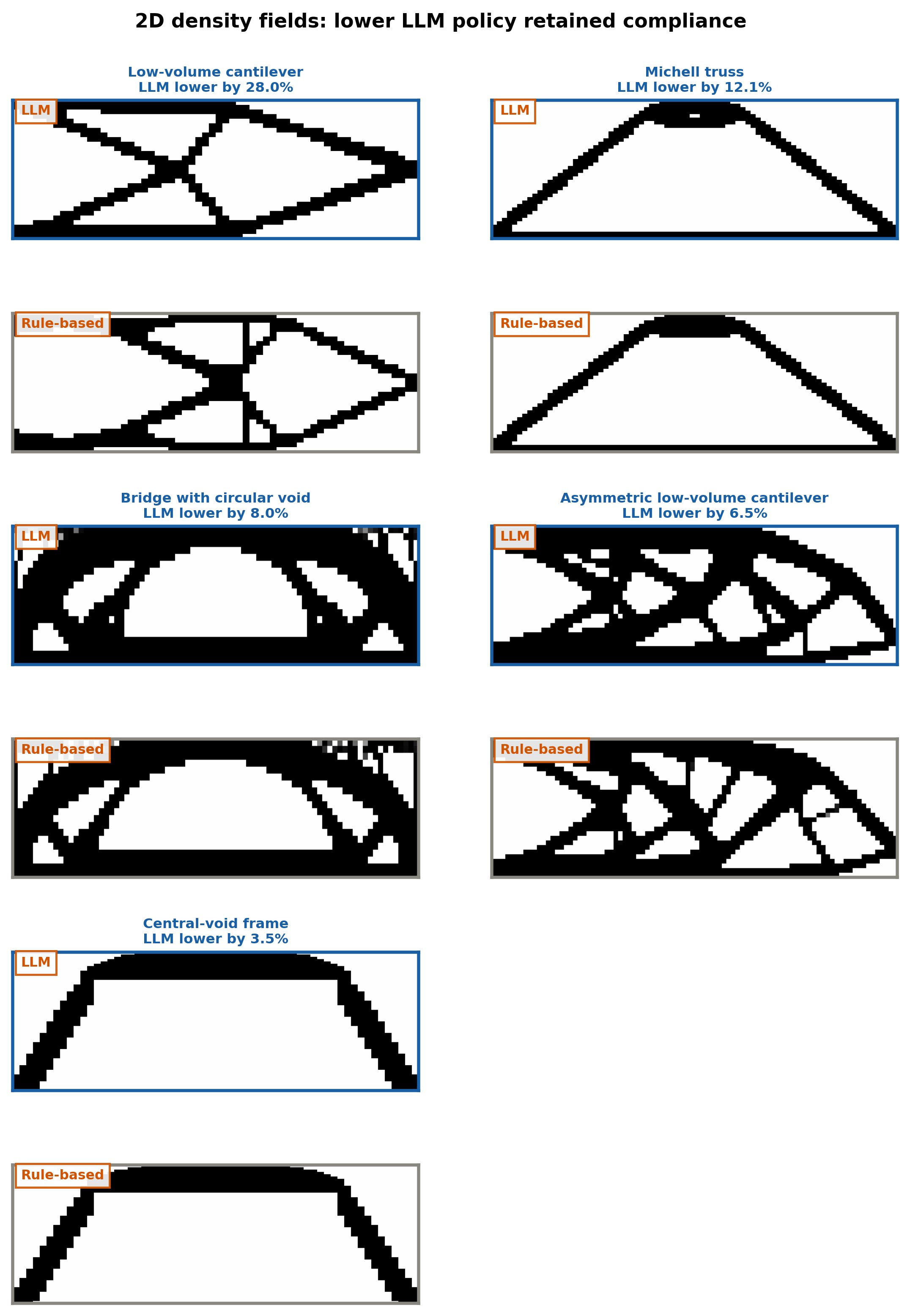}
  \caption{Representative seed-42 density fields for the five two-dimensional benchmark
    problems classified as lower soft-seed retained compliance after the 3\%/std guard
    ($\sigyield = 120.3$). Compliance labels and
    ordering use three-seed mean retained-best compliance $\Crep$. Each pair shows the LLM topology
(blue border, top) and rule-based topology (gray border, bottom).
    Benchmark label mappings are provided in Appendix~\ref{app:repro_map}.}
  \label{fig:all_topologies}
\end{figure}

\begin{figure}[p]
  \centering
  \includegraphics[width=0.92\linewidth]{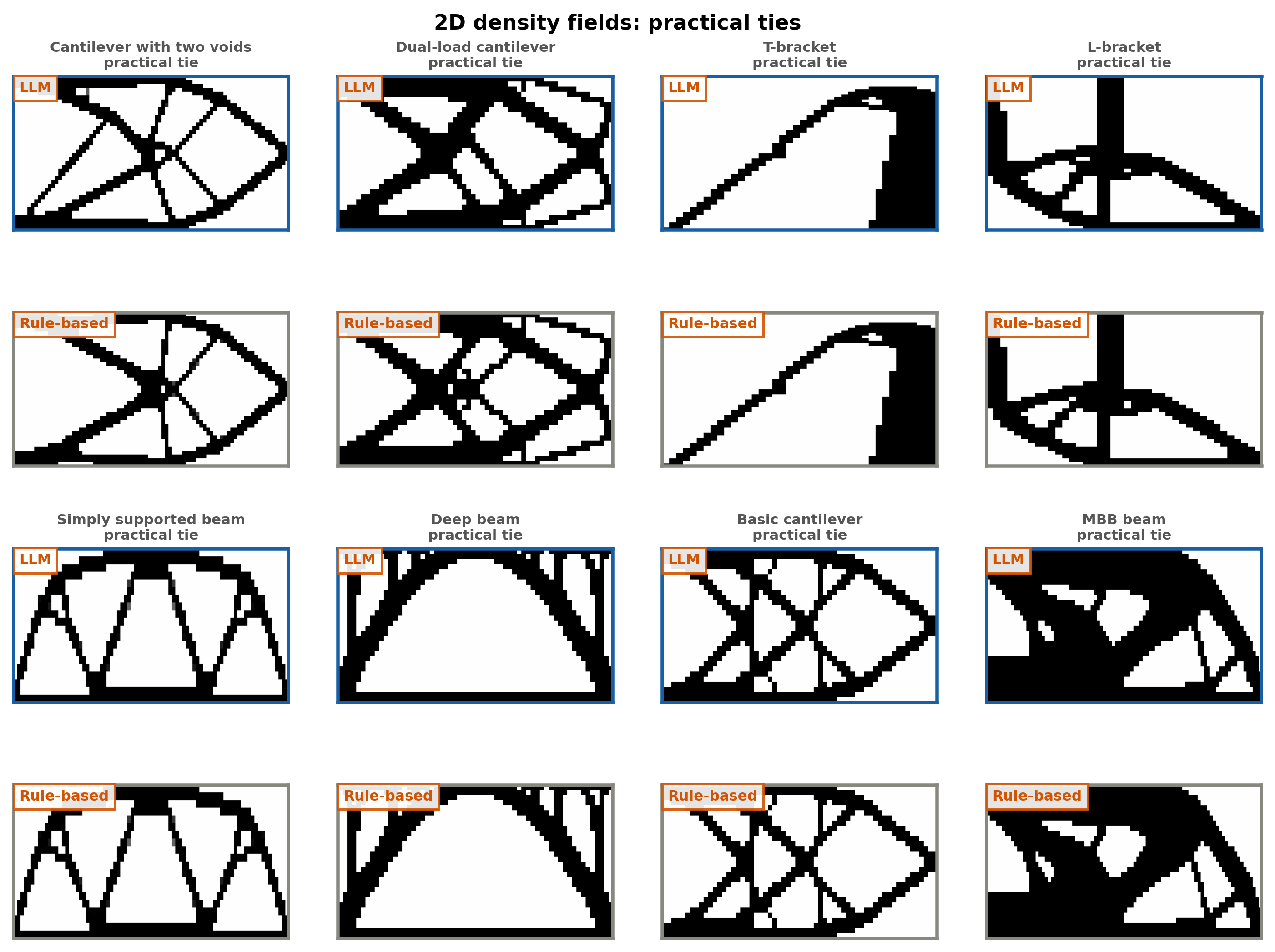}
  \caption{Representative seed-42 density fields for the two-dimensional benchmark cases that
    are practical ties under the 3\% tolerance and seed-variability guard
    ($\sigyield = 120.3$). Compliance labels and
    ordering use three-seed mean retained-best compliance $\Crep$.}
  \label{fig:all_topologies_ties}
\end{figure}

\begin{figure}[p]
  \centering
  \includegraphics[width=0.92\linewidth]{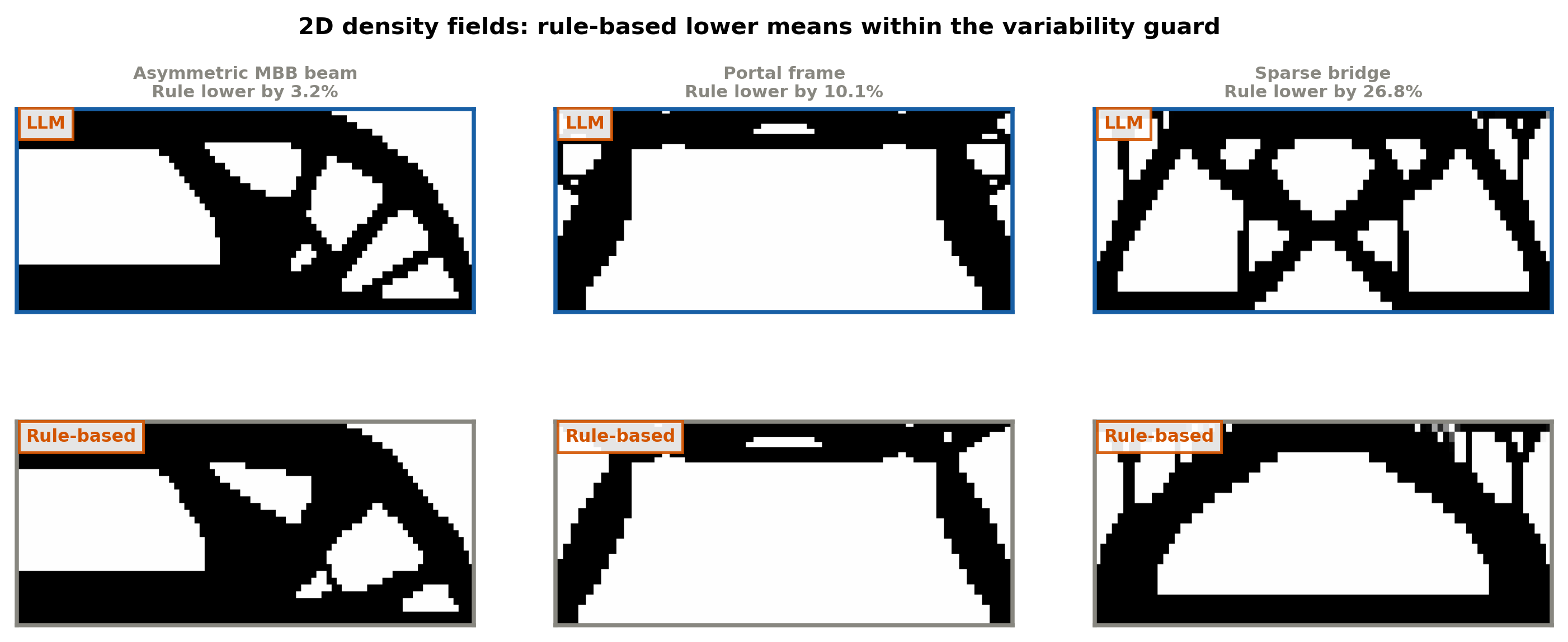}
  \caption{Representative seed-42 density fields for the three cases where
    the rule-based run has a lower retained compliance mean, but the difference
    is classified as a guarded tie in Table~\ref{tab:main_results}
    because it does not exceed the soft-seed variability guard. Compliance labels and ordering use three-seed mean
    retained-best compliance $\Crep$.}
  \label{fig:all_topologies_rule_favouring}
\end{figure}

Figures~\ref{fig:all_topologies}--\ref{fig:all_topologies_rule_favouring}
provide a visual overview of representative seed-42 density fields for all
16 problems.  The
largest retained-compliance soft-seed differences occur on the low-volume cantilever, Michell truss,
bridge with circular void, and asymmetric low-volume cantilever; several
difficult geometries are practical ties once a 3\% tolerance and seed
variability are applied.

\FloatBarrier

\subsection{Fixed-volume hotspot and input-ablation study}
\label{subsec:fixed_volume_study}

The fixed-volume study reruns the 16 two-dimensional problems with the initial
volume fraction held fixed throughout the controller loop.  It uses a separate
50th-percentile 2D stress calibration, giving $\sigyield=163.457$ for this
run.  The calibration population is different: the main $\sigyield=120.3$
threshold comes from the historical controller-policy traces whose controllers
could change volume fraction, whereas $\sigyield=163.457$ was recomputed for
the fixed-volume attribution reruns under the volume-lock protocol.  Because
this threshold differs from the original controller-policy comparison at
$\sigyield=120.3$, these results are not pooled
with Table~\ref{tab:main_results}.  They provide the most direct attribution
diagnostic for whether any apparent LLM benefit persists after
material-budget changes are disabled.

Tables~\ref{tab:fixed_volume_core} and
\ref{tab:fixed_volume_ablation_summary} score the fixed-volume policies by
feasibility-conditioned retained compliance, feasible-final compliance,
retained-any compliance, completed-evaluation counts, and trace availability.
The fixed-volume study weakens the causal interpretation of the controller-policy
retained-compliance difference.  The LLM condition completed 44 of 48 attempted
problem/repeat slots.  Four
attempted evaluations did not yield completed designs because execution ended
before evaluation; these are reported separately from completed but infeasible
designs.  The exact-hotspot condition completed 47 of 48 attempted
slots, and the rule, random, and input-ablation controls completed all 48.
Feasibility is therefore interpreted among completed evaluations, while missing
executions are reported separately rather than treated as structural
infeasibility.  Among completed evaluations, 25 of 44 soft-seed LLM records and
27 of 48 rule-based records produce an all-gate-passing retained state.  Seven
of the sixteen LLM/rule problem pairs are incomplete for
feasibility-conditioned pairwise comparisons.  Across the nine complete
LLM/rule paired problems, the LLM has lower mean $C_{\mathrm{feas}}$ in 5 cases,
rule-based control is lower in 0, and 4 are equal within table precision.
Feasible-final scoring is split: lower LLM 4, lower rule 4, equal 1.  The
corresponding geometric-mean ratios versus rule-based control are 0.995 for
feasible-retained compliance, 1.002 for feasible-final compliance, and 0.958
for retained-any compliance.  Thus the fixed-volume study controls the
material-budget degree of freedom but does not show a clear feasible-final LLM
performance advantage.

\begin{table}[htbp]
  \centering
  \footnotesize
  \caption{Fixed-volume 2D experiment scored by feasibility-conditioned retained compliance. Entries are means over feasible retained repeats. Parentheses report feasible/completed/attempted repeats, so attempted slots that ended before evaluation are separated from completed but infeasible designs. The final column reports the maximum absolute mean final-volume drift from the initial target across the displayed conditions.}
  \label{tab:fixed_volume_core}
  \begin{adjustbox}{max width=\linewidth}
  \begin{tabular}{lrrrrr}
    \toprule
    Problem & Soft-seed LLM & Exact hotspot & Rule-based controller & Random stress region & Max $|\overline{\Delta v_f}|$ \\
    \midrule
    Asymmetric low-volume cantilever & -- (0/3/3) & -- (0/3/3) & -- (0/3/3) & -- (0/3/3) & 0.000 \\
    Sparse bridge & -- (0/3/3) & -- (0/3/3) & -- (0/3/3) & -- (0/3/3) & 0.000 \\
    Bridge with circular void & -- (0/3/3) & -- (0/3/3) & -- (0/3/3) & -- (0/3/3) & 0.000 \\
    Low-volume cantilever & -- (0/3/3) & -- (0/3/3) & -- (0/3/3) & -- (0/3/3) & 0.000 \\
    Basic cantilever & 94.75 (3/3/3) & 93.87 (3/3/3) & 94.75 (3/3/3) & 94.75 (3/3/3) & 0.000 \\
    Cantilever with two voids & 134.64 (3/3/3) & 135.01 (3/3/3) & 135.94 (3/3/3) & 135.01 (3/3/3) & 0.000 \\
    Deep beam & 34.17 (3/3/3) & 35.08 (3/3/3) & 35.08 (3/3/3) & 35.05 (3/3/3) & 0.000 \\
    Dual-load cantilever & -- (0/1/3) & -- (0/3/3) & -- (0/3/3) & -- (0/3/3) & 0.000 \\
    Central-void frame & 18.19 (3/3/3) & 18.20 (3/3/3) & 18.20 (3/3/3) & 18.20 (3/3/3) & 0.000 \\
    L-bracket & 93.75 (3/3/3) & 93.75 (3/3/3) & 93.75 (3/3/3) & 93.75 (3/3/3) & 0.000 \\
    Asymmetric MBB & -- (0/3/3) & -- (0/3/3) & -- (0/3/3) & -- (0/3/3) & 0.000 \\
    MBB beam & -- (0/3/3) & -- (0/3/3) & -- (0/3/3) & -- (0/3/3) & 0.000 \\
    Michell truss & 37.96 (3/3/3) & 38.38 (2/2/3) & 38.38 (3/3/3) & 37.57 (3/3/3) & 0.000 \\
    Portal frame & 98.57 (3/3/3) & 98.57 (3/3/3) & 98.57 (3/3/3) & 98.57 (3/3/3) & 0.000 \\
    Simply supported & 66.26 (3/3/3) & 66.26 (3/3/3) & 66.26 (3/3/3) & 66.26 (3/3/3) & 0.000 \\
    T-bracket & 11.96 (1/1/3) & 11.96 (3/3/3) & 11.97 (3/3/3) & 11.97 (3/3/3) & 0.000 \\
    \bottomrule
  \end{tabular}
  \end{adjustbox}
\end{table}

\begin{table}[htbp]
  \centering
  \footnotesize
  \caption{Fixed-volume 2D ablation summary. Completion reports completed evaluations out of attempted condition/problem/repeat slots. Feasibility counts are feasible/completed evaluations, not feasible/all-attempted slots; missing executions are reported separately. Ratios are geometric means versus fixed-volume rule-based control over complete problem pairs. Trace/action counts used for localization diagnostics are reported in Appendix~\ref{app:repro_map}.}
  \label{tab:fixed_volume_ablation_summary}
  \begin{adjustbox}{max width=\linewidth}
  \begin{tabular}{lrrrrrrrr}
    \toprule
    Condition & Completed & Feas. retained & Feas. final & Missing exec. & $C_{\mathrm{feas}}/$rule & $C_{\mathrm{final,feas}}/$rule & $C_{\mathrm{any}}/$rule & Paired problems \\
    \midrule
    Soft-seed LLM & 44/48 & 25/44 & 25/44 & 4 & 0.995 & 1.002 & 0.958 & 9/16 \\
    Exact hotspot & 47/48 & 26/47 & 26/47 & 1 & 0.998 & 1.000 & 1.000 & 9/16 \\
    Rule-based controller & 48/48 & 27/48 & 27/48 & 0 & 1.000 & 1.000 & 1.000 & 9/16 \\
    Random stress region & 48/48 & 27/48 & 27/48 & 0 & 0.997 & 1.000 & 0.967 & 9/16 \\
    Density-only LLM & 48/48 & 27/48 & 27/48 & 0 & 0.998 & 1.001 & 0.787 & 9/16 \\
    Stress-only LLM & 48/48 & 27/48 & 27/48 & 0 & 0.996 & 0.997 & 0.798 & 9/16 \\
    Numeric-only LLM & 48/48 & 27/48 & 27/48 & 0 & 0.994 & 0.995 & 0.828 & 9/16 \\
    Global-only LLM & 48/48 & 27/48 & 27/48 & 0 & 1.000 & 1.000 & 1.000 & 9/16 \\
    \bottomrule
  \end{tabular}
  \end{adjustbox}
\end{table}

\begin{figure}[!htbp]
  \centering
  \includegraphics[width=0.92\textwidth,height=0.82\textheight,keepaspectratio]{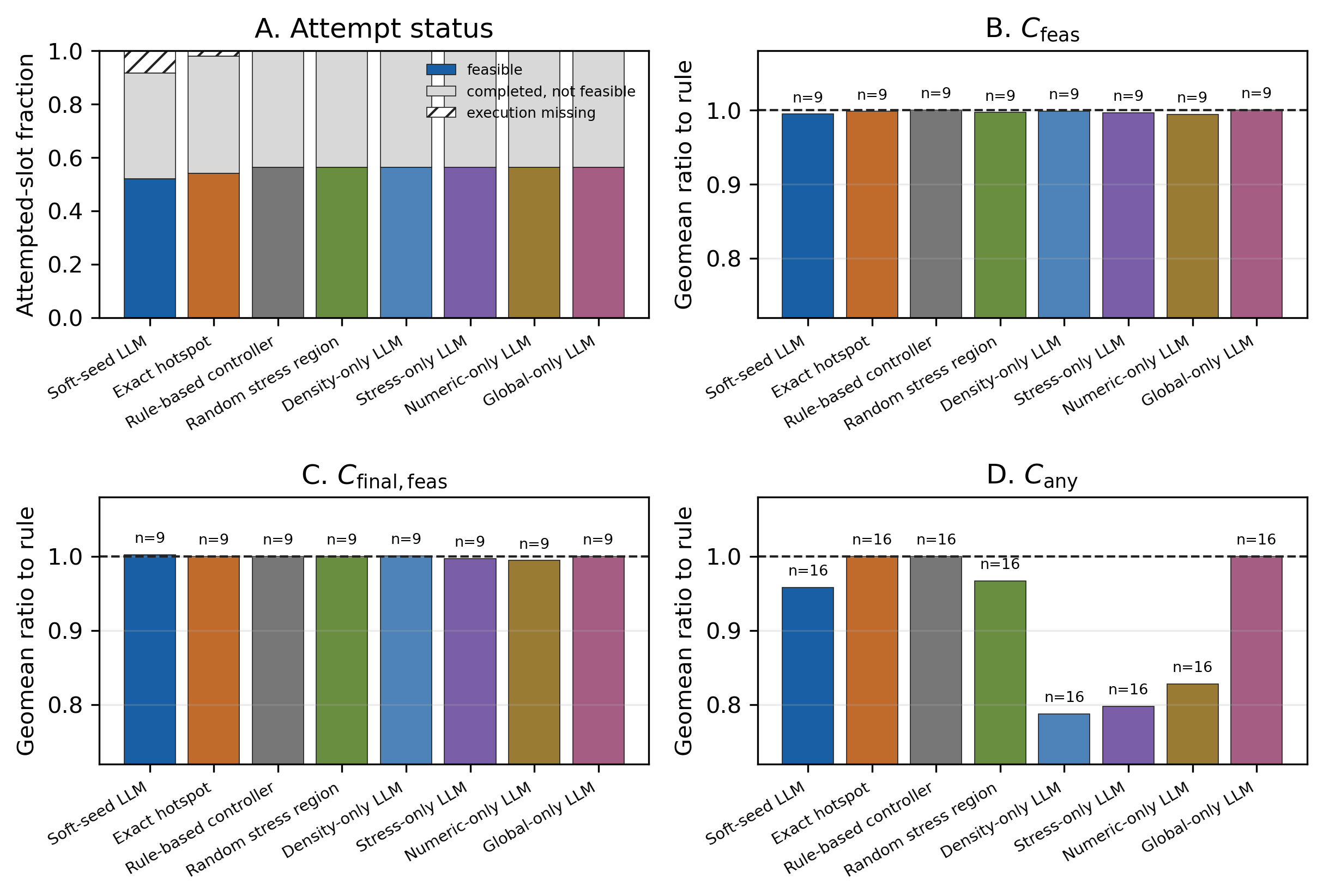}
  \caption{Fixed-volume 2D primary endpoints. Panel A decomposes attempted
    condition/problem/repeat slots into feasible retained designs, completed
    but non-feasible evaluations, and missing executions. Panels B--D report
    geometric-mean ratios to fixed-volume rule-based control for
    feasibility-conditioned retained compliance, feasible-final compliance,
    and retained-any compliance, respectively; $n$ gives the number of complete
    problem pairs used for each geometric mean. Values below one have lower
    compliance than rule-based control.}
  \label{fig:fixed_volume_feasible}
\end{figure}

\FloatBarrier
The deterministic max-stress hotspot baseline is competitive with the LLM
policy under the feasible-primary endpoints.  Against exact hotspot seeding,
the LLM has lower feasible-retained compliance in 5 of 9 complete paired
problems, exact hotspot is lower in 1, and 3 are equal.  Under feasible-final
scoring, exact hotspot is lower in 5 of the 9 complete paired problems, while
the LLM is lower in 4.
This comparison addresses the central attribution question: the present data
do not show that a vision LLM locates stress interventions better than a simple
algorithm using the numerical stress field.

Direct localization metrics lead to the same conclusion, with an important
trace-coverage limitation.  The recovered seed-7 scalar summary contributes to
the completed-evaluation counts above, but its per-step traces were not
available for localization/action diagnostics; the LLM localization analysis
therefore covers 28 completed evaluations with per-step records.  Over 76 accepted
LLM spatial actions in those traces, the mean normalized distance between the
LLM seed center and the true maximum-stress element is 0.221 domain diagonals.
Mean overlap of the seed region with the top 1\%, 5\%, and 10\% stress regions
is 0.041, 0.118, and 0.196, respectively.  These values show that some LLM
proposals intersect high-stress regions, but the localization is not precise
enough to establish reliable hotspot targeting.  The density-only, stress-only, numeric-only, and
global-action-only ablations further show that retained-any compliance can
improve without producing a comparable feasible-final advantage.  The
fixed-volume study therefore improves attribution clarity while weakening any
claim that LLM visual reasoning is the causal source of the observed
controller-policy differences.

\begin{figure}[!htbp]
  \centering
  \includegraphics[width=0.92\textwidth,height=0.82\textheight,keepaspectratio]{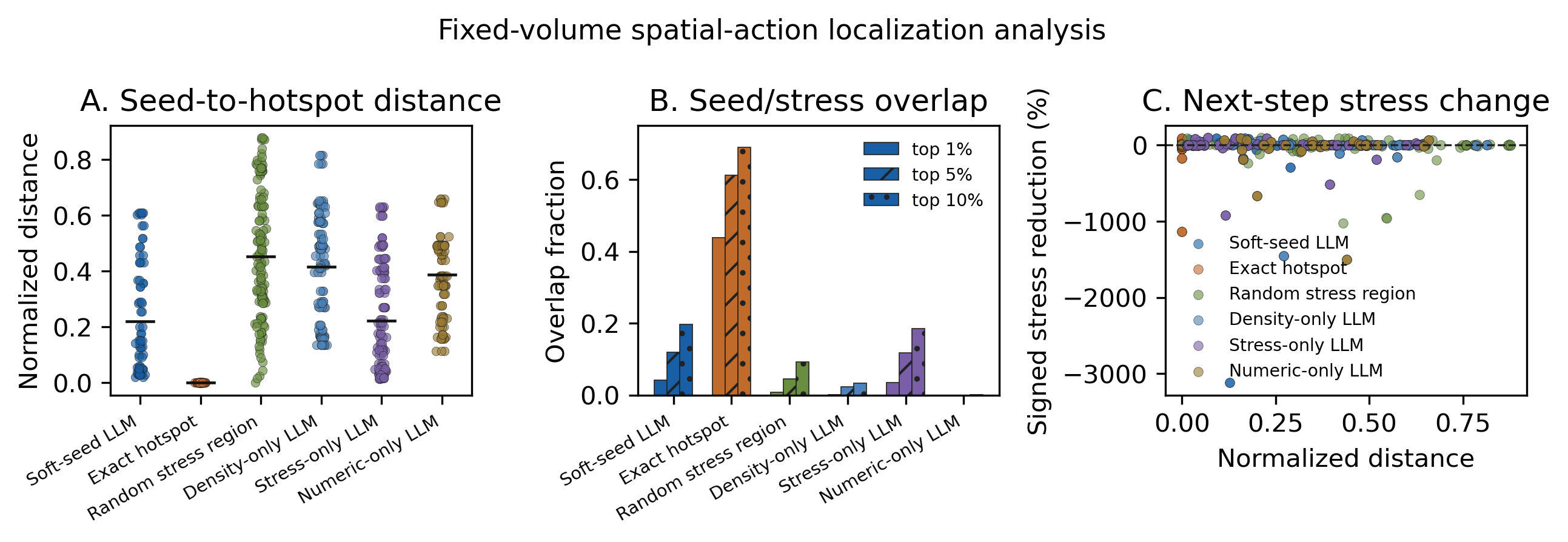}
  \caption{Spatial-action localization analysis across fixed-volume policies.
    Panel A shows per-action normalized seed-to-hotspot distances with mean
    bars. Panel B reports seed overlap with the top 1\%, 5\%, and 10\%
    von~Mises stress regions. Panel C couples the accepted spatial action to
    signed next-step stress reduction,
    $100(\sigma_{\mathrm{current}}-\sigma_{\mathrm{next}})/\sigma_{\mathrm{current}}$;
    positive values indicate stress reduction and negative values indicate
    stress increase. Exact hotspot seeding is the zero-distance reference. The figure uses only
    completed evaluations with available per-step traces.}
  \label{fig:fixed_volume_localization}
\end{figure}

\clearpage

\subsection{Sensitivity to \texorpdfstring{$\sigyield$}{sigma-yield}}

\begin{figure}[htbp]
  \centering
  \includegraphics[width=0.8\linewidth]{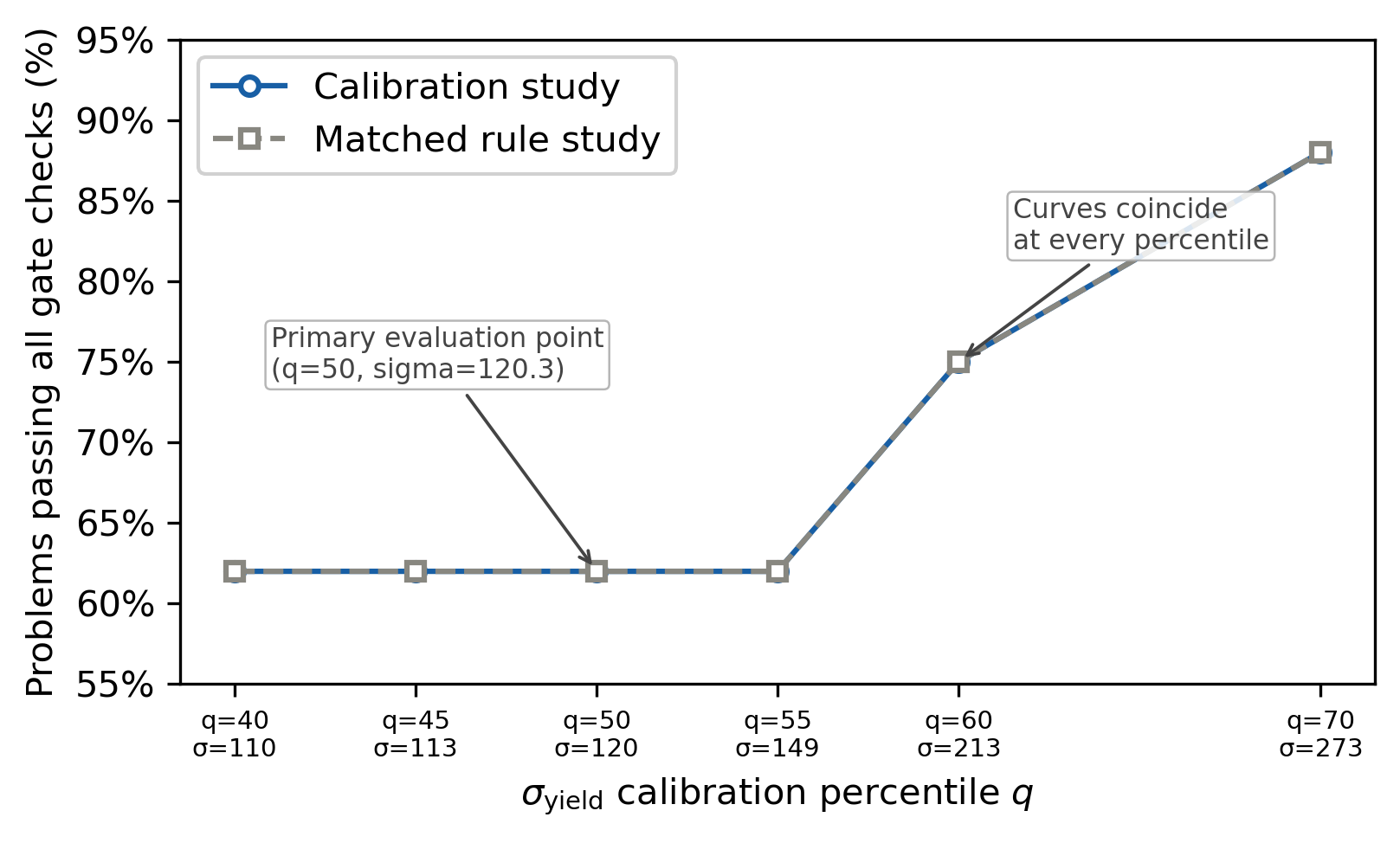}
  \caption{Stress-satisfaction rate vs.\ $\sigyield$ calibration
    percentile. The deterministic calibration study uses the rule-based
    interpreter path and therefore coincides with the matched rule study. This figure is
    calibration evidence, not controller-comparison evidence.}
  \label{fig:sensitivity}
\end{figure}

The primary evaluation above is conducted at a single calibration point.
Figure~\ref{fig:sensitivity} and Table~\ref{tab:sensitivity} report the
stress-gate pass count across six calibration levels.  These single-seed sweeps are
used as calibration evidence only: the deterministic calibration study uses
the rule-based interpreter path, so it coincides with the matched rule study.
The gate-pass count is 10/16 for both studies from calibration percentile $q=40$
through $q=55$, rises to 12/16 at $q=60$, and reaches 14/16 at $q=70$ as the stress
gate becomes loose.

\begin{table}[!htbp]
  \caption{Sensitivity of stress-satisfaction rate to $\sigyield$.
    \% fail at initial solve: fraction failing the stress gate on the
    compliance-only solve. ``Gate pass'' means all seven evaluator gate
    checks pass; four diagnostics are recorded but do not gate validity.
    Single run per condition per calibration level; $N=3$ repeated runs are
    reported only at $q=50$ (primary).
    Boldface marks the primary evaluation point.}
  \label{tab:sensitivity}
  \centering
  \begin{tabular}{rrrrrr}
    \toprule
    $q$ & $\sigyield$ & \% fail initially & Calibration study & Matched rule study & Outcome \\
    \midrule
    40  & 110.4 & 56\% & 10/16 (62.5\%) & 10/16 (62.5\%) & Same  \\
    45  & 112.8 & 56\% & 10/16 (62.5\%) & 10/16 (62.5\%) & Same  \\
    \textbf{50} & \textbf{120.3} & \textbf{50\%} &
      \textbf{10/16 (62.5\%)} & \textbf{10/16 (62.5\%)} & \textbf{Same} \\
    55  & 148.5 & 44\% & 10/16 (62.5\%) & 10/16 (62.5\%) & Same  \\
    60  & 213.3 & 38\% & 12/16 (75\%) & 12/16 (75\%) & Same \\
    70  & 272.7 & 31\% & 14/16 (87.5\%) & 14/16 (87.5\%) & Same  \\
    \bottomrule
  \end{tabular}
\end{table}

\begin{figure}[!htbp]
  \centering
  \includegraphics[width=\linewidth]{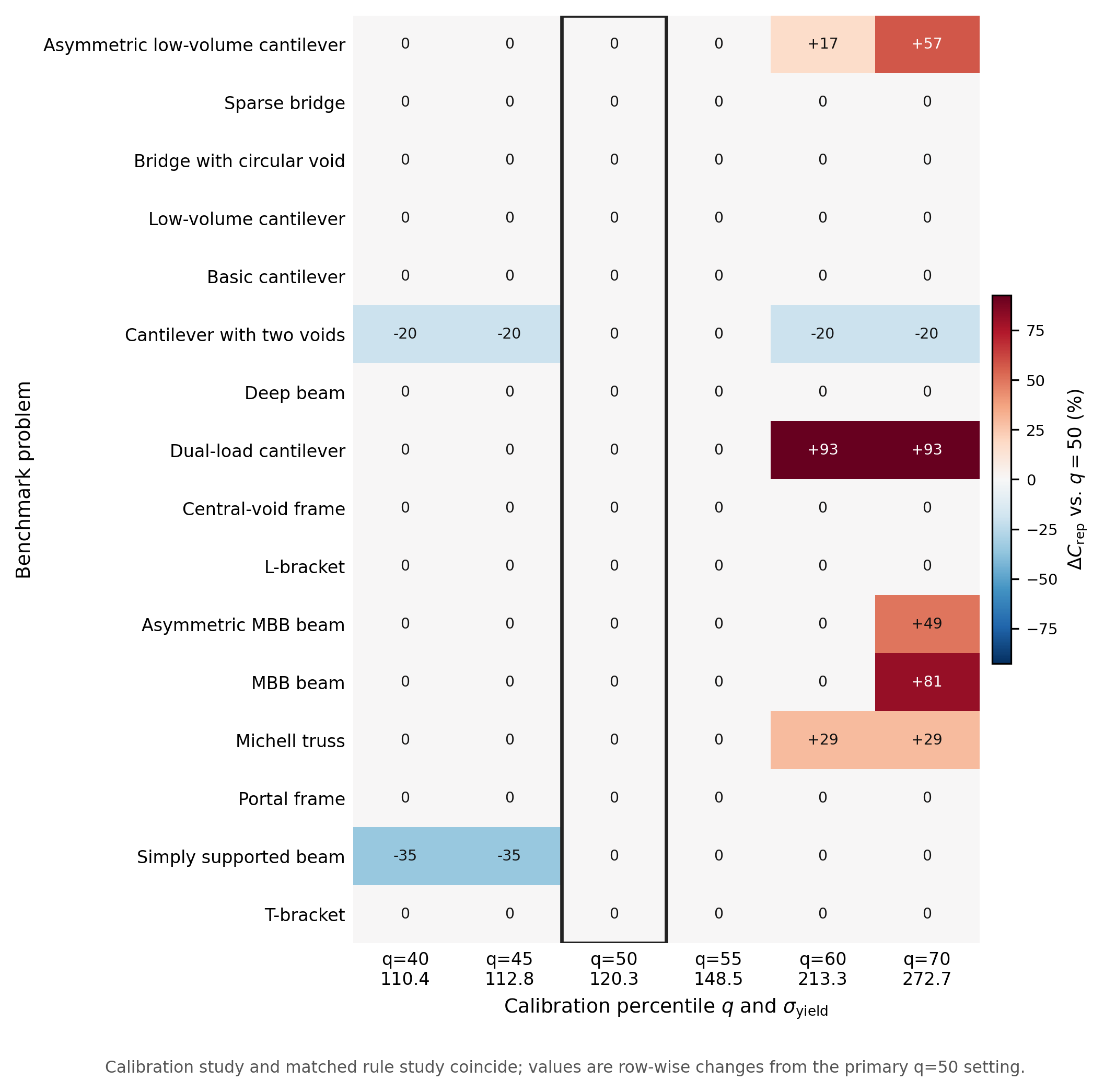}
  \caption{Per-problem retained-compliance sensitivity to the
    $\sigyield$ calibration level. Cell values show the percentage change in
    retained-best compliance $\Crep$ relative to the primary $q=50$ value for
    the same problem; negative values indicate lower retained compliance than
    at $q=50$. The calibration study uses the rule-based interpreter path, so
    this is calibration evidence rather than an LLM-vs-rule comparison.
    Problems are sorted by their primary-evaluation ordering, and the boxed
    column marks $q=50$, $\sigyield = 120.3$.}
  \label{fig:cross_pct}
\end{figure}

Figure~\ref{fig:cross_pct} disaggregates the calibration sensitivity analysis to show
how each problem's retained compliance changes relative to the primary
$q=50$ calibration level.  Because the sweep is not an LLM-controller run, it
should be read as a map of stress-threshold sensitivity rather than evidence
for controller benefit.

\FloatBarrier

\subsection{Convergence analysis}

\begin{figure}[H]
  \centering
  \includegraphics[width=\linewidth]{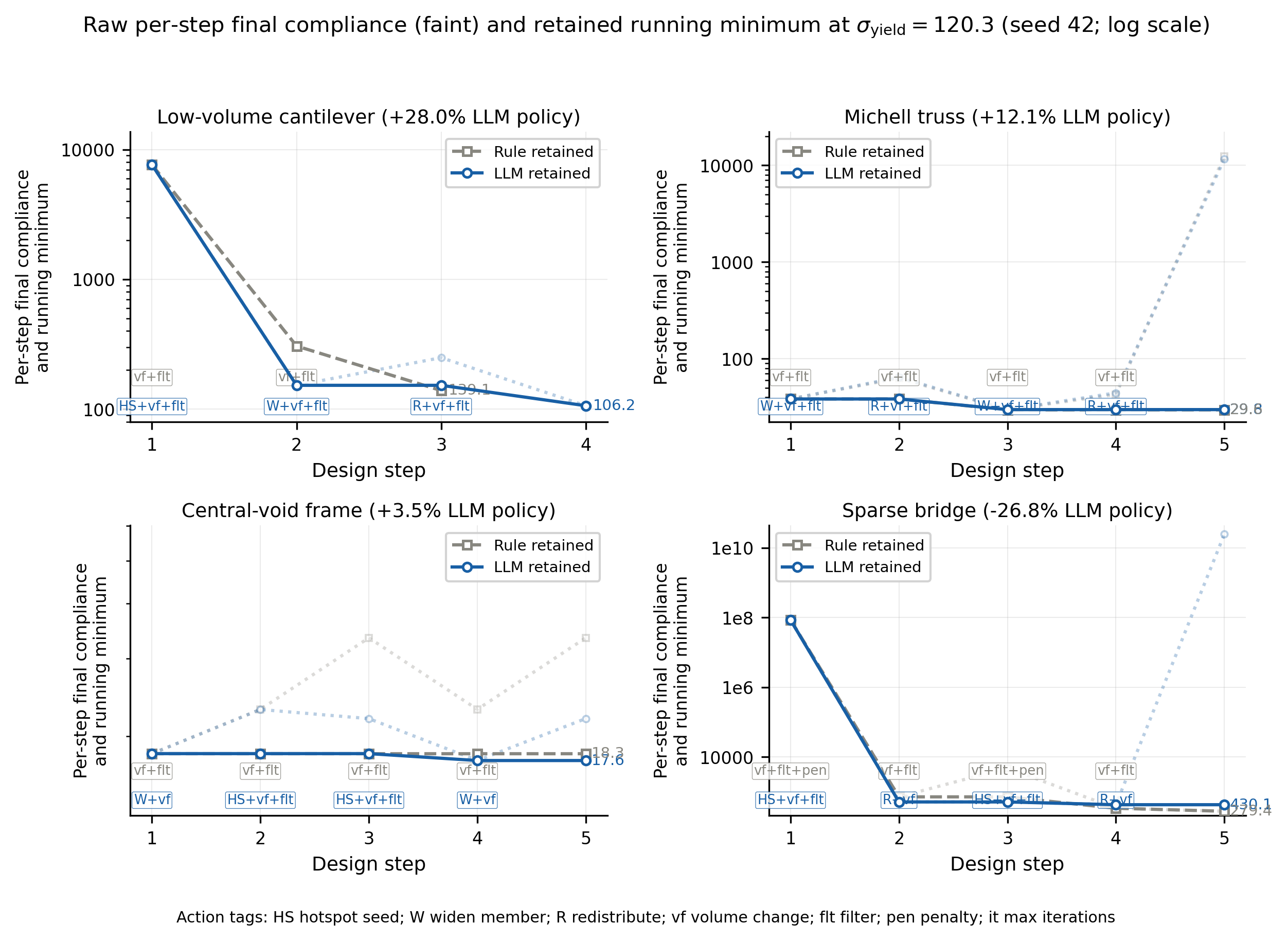}
  \caption{Step-by-step compliance convergence for seed-42 traces
    ($\sigyield = 120.3$).
    Faint lines show raw per-step final compliance; bold lines show the
    retained running minimum used by the trace-level selection rule.  The
    small action tags identify applied controller actions
    (HS=hotspot seed, W=widen member, R=redistribute, vf=volume fraction,
    flt=filter radius, pen=penalization). The logarithmic y-axis keeps
    exploratory spikes visible.  Scalar tables instead report retained-best
    compliance aggregated over three seeds.}
  \label{fig:convergence}
\end{figure}

Figure~\ref{fig:convergence} shows step-by-step compliance for four
representative seed-42 traces.

\FloatBarrier

\begin{figure}[H]
  \centering
  \begin{subfigure}[t]{\linewidth}
    \makebox[\linewidth][c]{\includegraphics[width=1.08\linewidth]{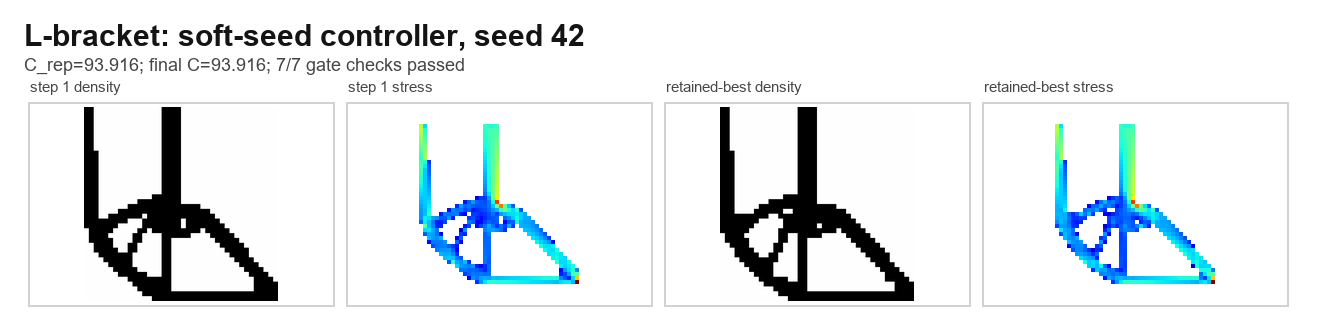}}
    \caption{Soft-seed LLM condition: step~1 and retained-best density/stress
      fields for the L-bracket controller sequence.  The final state passes all
      seven gate checks, with $\Crep = 93.916$.}
  \end{subfigure}
  \vspace{4pt}
  \begin{subfigure}[t]{\linewidth}
    \makebox[\linewidth][c]{\includegraphics[width=1.08\linewidth]{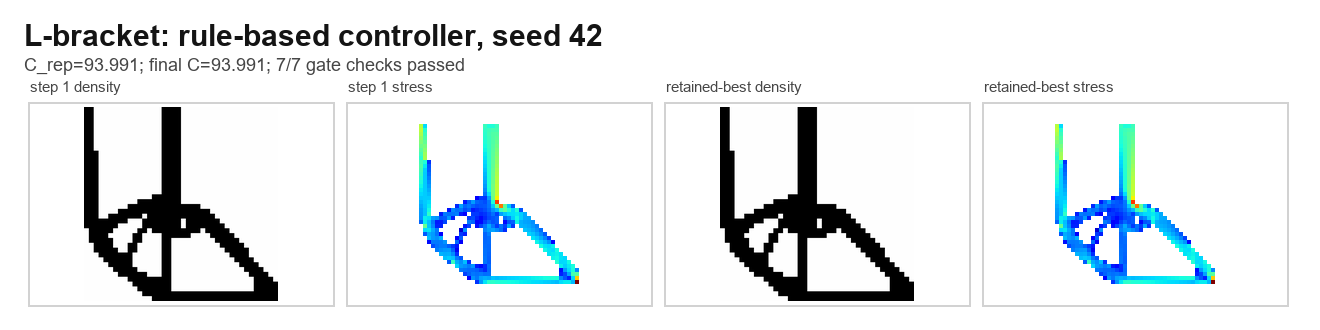}}
    \caption{Rule-based condition: step~1 and retained-best density/stress
      fields for the matched L-bracket run.  The final state passes all
      seven gate checks, with $\Crep = 93.991$.}
  \end{subfigure}
  \caption{Controller sequence for the L-bracket in the primary seed-42 evaluation run
    ($\sigyield = 120.3$, seed~42).  The panels show step~1 density,
    step~1 von~Mises stress, retained-best density, and retained-best
    von~Mises stress.}
  \label{fig:step_gallery}
\end{figure}

Figure~\ref{fig:step_gallery} shows the controller sequence for a
representative L-bracket repeat with seed 42.  Under this trace, both the soft-seed LLM
condition and the rule-based condition pass all seven gate checks.  The figure is
therefore used as a controller-sequence illustration rather than as evidence of a practical
retained-compliance advantage: the density and stress fields show that the LLM action only
slightly changes the retained L-bracket topology relative to the rule-based
condition.

\begin{figure}[H]
  \centering
  \includegraphics[width=0.92\linewidth]{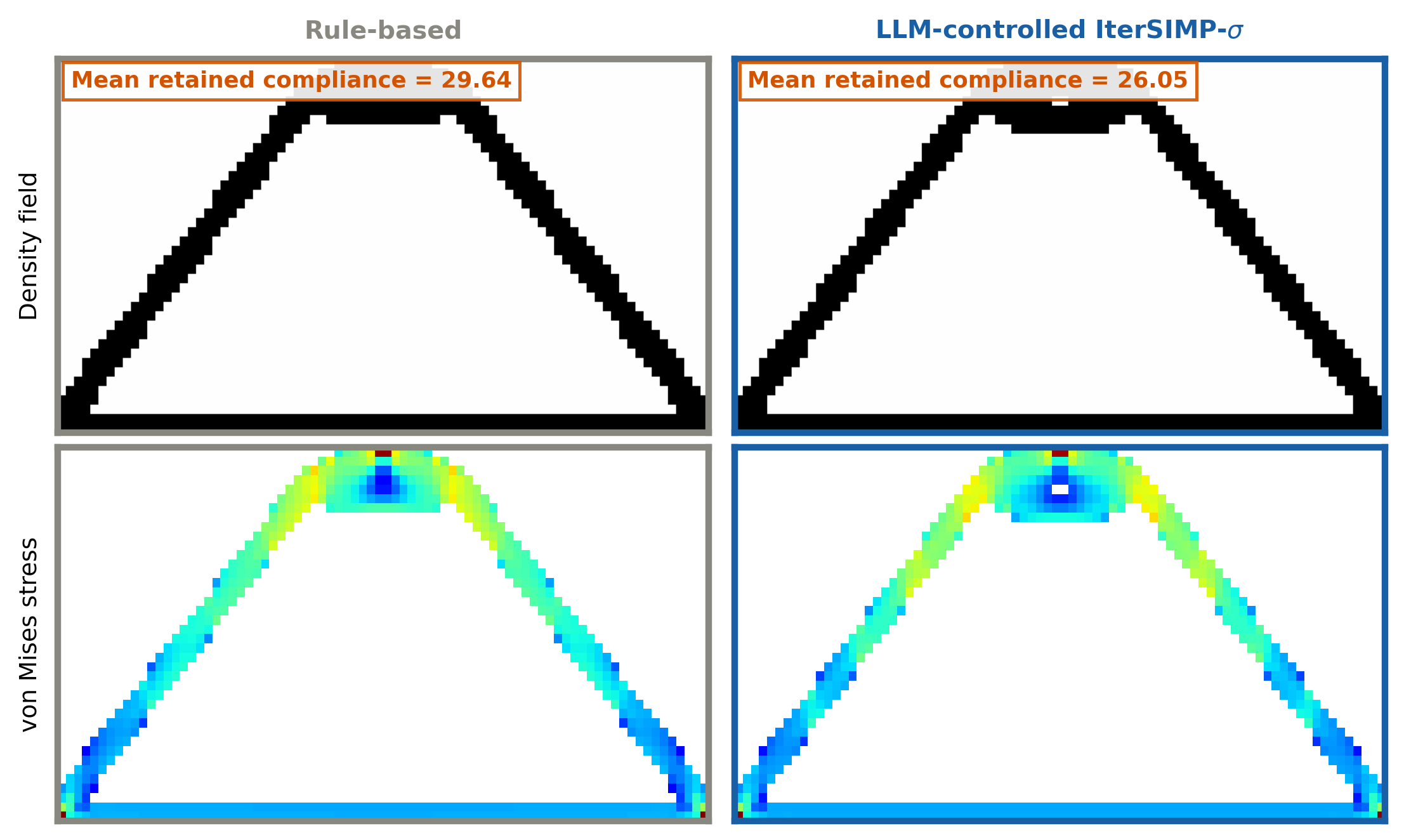}
  \caption{Representative seed-42 density and von~Mises stress fields for the Michell truss
    ($\sigyield = 120.3$). Labels report
    three-seed mean retained-best compliance $\Crep$: rule-based (left, $C=29.64$) vs.\ soft-seed
LLM (right, $C=26.05$, $12.1\%$ lower retained compliance). Stress colors are normalized per
    panel.}
  \label{fig:case_study}
\end{figure}

Figure~\ref{fig:case_study} provides the analogous density-and-stress
comparison for the Michell truss, the most geometrically nuanced
problem in the suite.
The three-seed soft-seed mean is lower than rule-based compliance
($26.05$ vs.\ $29.64$), although seed-level behavior remains variable.
The visual comparison shows the geometry and stress-field differences behind
that aggregate result.

\FloatBarrier

\subsection{Variance and high-CV problems}
\label{subsec:variance}

\begin{figure}[!htbp]
  \centering
  \includegraphics[width=\linewidth]{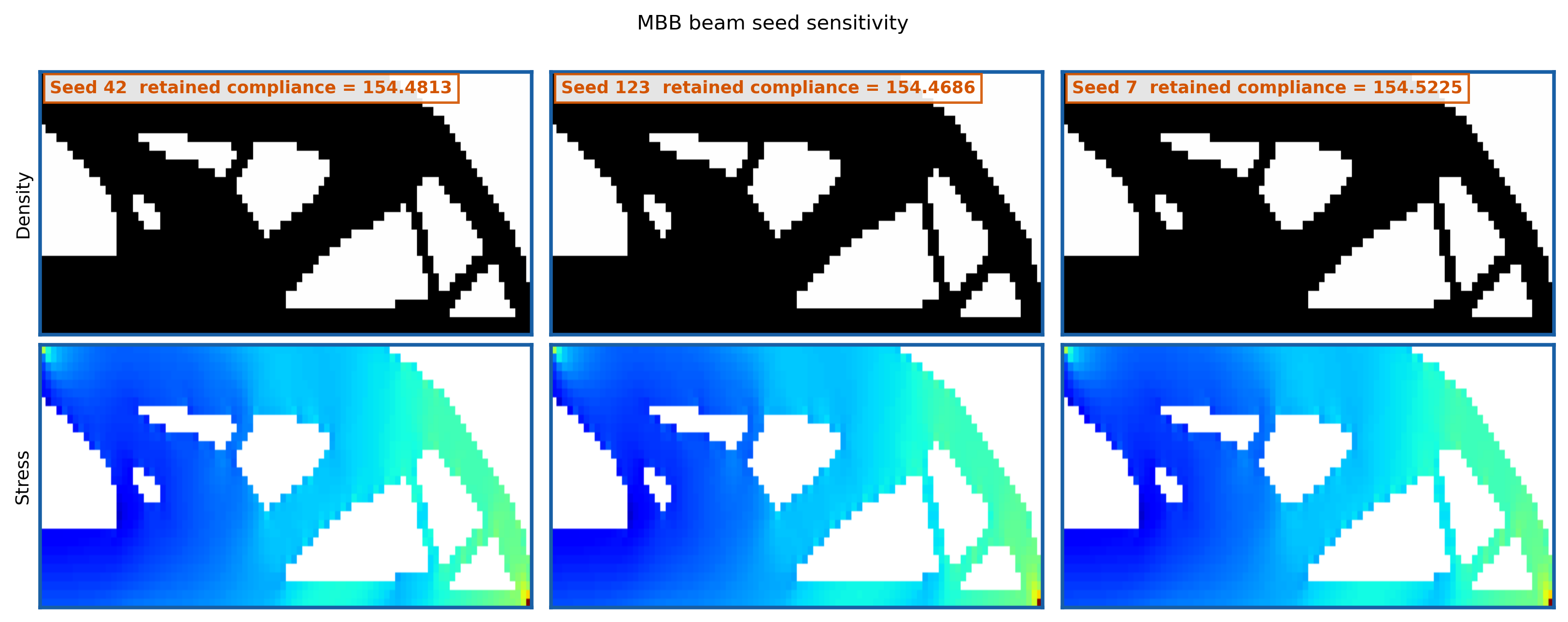}
  \caption{\textbf{MBB beam} seed sensitivity (CV~$=0.01\%$).
    All three soft-seed repeats are nearly identical
    ($C=154.48$, $154.47$, $154.52$), producing slightly lower rule-based
    retained compliance rather than a soft-seed advantage.}
  \label{fig:seed_mbb}
\end{figure}

\begin{figure}[!htbp]
  \centering
  \includegraphics[width=\linewidth]{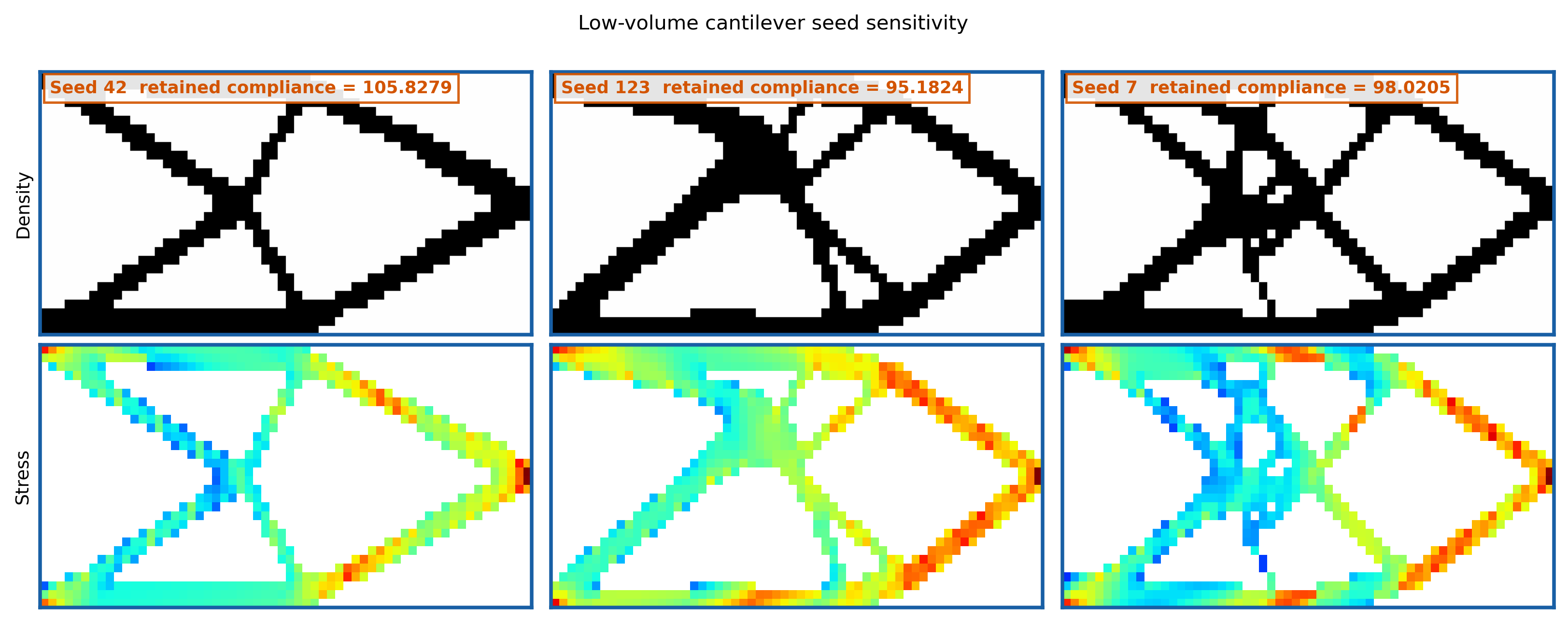}
  \caption{\textbf{Low-volume cantilever} seed sensitivity (CV~$=4.5\%$).
    The three soft-seed repeats pass all seven gate checks with
    $C=105.83$, $95.18$, and $98.02$.}
  \label{fig:seed_cant15}
\end{figure}

\begin{figure}[!htbp]
  \centering
  \includegraphics[width=\linewidth]{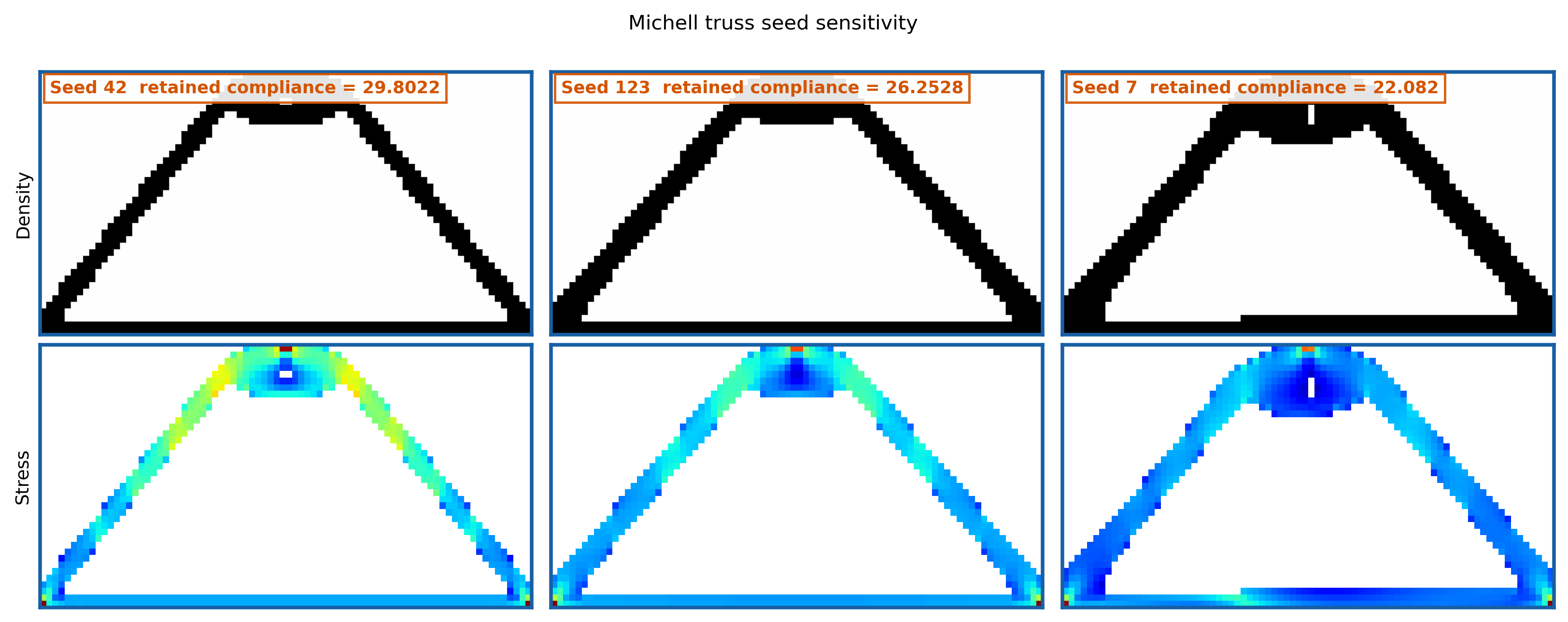}
  \caption{\textbf{Michell truss} seed sensitivity (CV~$=12.1\%$).
    The three soft-seed repeats give $C=29.80$, $26.25$, and $22.08$;
    the first final state fails gate checks and the latter two pass all seven
    gate checks.
    Figures~\ref{fig:seed_mbb}--\ref{fig:seed_michell} together
    constitute the seed sensitivity analysis (LLM condition; density
    top row, von~Mises stress bottom row, seeds~42/123/7 left to right).}
  \label{fig:seed_michell}
\end{figure}

Rule-based compliance is nearly deterministic across seeds
(median CV~$= 0.0\%$), reflecting the absence of stochastic elements
in the rule-based action sequence once the SIMP solver follows a deterministic
initialization and controller state.
Soft-seed compliance is stable on 11 of 16 problems (CV~$< 5\%$).
The largest observed variability occurs on the sparse bridge (CV~=~23.9\%),
Michell truss (12.1\%), portal frame (12.0\%),
asymmetric low-volume cantilever (6.4\%), and bridge with circular void (5.3\%).
Figures~\ref{fig:seed_mbb}--\ref{fig:seed_michell} provide seed-level
visual comparisons for representative cases in the figure set.

\textbf{MBB beam} (CV~$=0.01\%$, Figure~\ref{fig:seed_mbb}):
all three soft-seed repeats converge to nearly identical compliance
($C=154.48$, $154.47$, $154.52$).  The soft-seed condition is therefore a
small rule-based loss on this problem rather than a bimodal success case.

\textbf{Low-volume cantilever} (CV~$=4.5\%$, Figure~\ref{fig:seed_cant15}):
all three repeats pass all seven gate checks, with compliance values
$C=105.83$, $95.18$, and $98.02$.  The spread reflects modest
initialization sensitivity within the same broad X-truss basin.

\textbf{Michell truss} (CV~$=12.1\%$, Figure~\ref{fig:seed_michell}):
the three repeats give $C=29.80$, $26.25$, and $22.08$; the first final state
fails gate checks and the latter two pass all seven gate checks.  The mean is
lower than rule-based compliance, but the seed-level feasibility and
compliance variation should be treated as material uncertainty.

\FloatBarrier
\subsection{3D benchmark}
\label{subsec:3d}

Table~\ref{tab:3d} reports the exploratory controller-policy results for the six
three-dimensional problems under the same paper-wide threshold
$\sigyield = 120.3$.  Both conditions pass all seven gate checks on all six
reported summaries.  In this exploratory subset, the LLM controller-policy
condition has lower retained compliance on 4 of 6 problems and ties the
remaining 2, but these differences should not be read as fixed-volume
compliance reductions:
the controller changes the volume-fraction target on several 3D traces.
The bridge row uses a validity-filtered bridge evaluation for the same benchmark setting satisfying the
predefined finite-positive-compliance and seven-gate validity criteria in
Appendix~\ref{app:repro_map}.  Because this 3D subset does not include fixed-volume
attribution studies or a no-seed 3D ablation, it is controller-level evidence rather
than an isolated estimate of the seed materialization effect.

\begin{table}[!htbp]
  \caption{3D benchmark: LLM vs.\ rule-based, $\sigyield = 120.3$.
    Entries are three-seed means of retained-best compliance
    $\Crep$; descriptive population standard deviations are zero for these
    3D scalar summaries and are omitted.  Final vf gives the final
    volume-fraction target for LLM/rule traces.
    The three 3D scalar repeats were identical under the recorded
    deterministic controller/solver paths, so these repeats should be read as
    trace reproducibility checks rather than independent stochastic samples.
    Both conditions pass all seven gate checks on all 6 reported summaries.
    $\Delta$: relative retained-compliance difference
    $= (C_{\text{rule}} - C_{\text{LLM}})/C_{\text{rule}} \times 100\%$.
    The 3D bridge LLM value and visual panels use the validity-filtered bridge
    evaluation checked in Appendix~\ref{app:repro_map}; the original run
    failed the predefined finite-positive-compliance validity criterion.
    The 3D bridge rule value uses the lower final state because that logged
    endpoint satisfies the same finite-positive, grayness, and seven-gate
    validity checks as the retained tracker.
    This is not a fixed-volume comparison and should not be interpreted as a
    validated 3D performance claim. Lower retained compliance under the LLM
    policy: 4/6; ties: 2/6.}
  \label{tab:3d}
  \centering
  \small
  \begin{adjustbox}{width=\linewidth}
  \begin{tabular}{lrrrrrrrrr}
    \toprule
    Problem & Mesh & $n_e$
      & \multicolumn{2}{c}{$\Crep$}
      & $\Delta$~(\%)
      & \multicolumn{2}{c}{Steps}
      & Final vf
      & Class \\
    \cmidrule(lr){4-5}\cmidrule(lr){7-8}
    & & & LLM & Rule & & LLM & Rule & L/R & \\
    \midrule
    3D cantilever & $40{\times}20{\times}10$ & 8000
      &  4.884 &  5.916 & $+17.4$ & 2 & 1 & 0.31/0.25 & \textbf{LLM} \\
    3D MBB beam        & $40{\times}14{\times}6$  & 3360
      & 30.431 & 38.349 & $+20.6$ & 2 & 1 & 0.24/0.18 & \textbf{LLM} \\
    3D bridge     & $30{\times}10{\times}6$  & 1800
      &  3.058 &  3.201 & $+4.4$ & 5 & 2 & 0.32/0.26 & \textbf{LLM} \\
    3D L-bracket & $20{\times}20{\times}5$  & 2000
      & 11.859 & 11.859 & $\phantom{-}0.0$ & 1 & 1 & 0.25/0.25 & Tie \\
    3D Michell truss    & $24{\times}12{\times}6$  & 1728
      &  3.341 &  4.400 & $+24.1$ & 2 & 1 & 0.18/0.12 & \textbf{LLM} \\
    3D torsion block    & $20{\times}10{\times}8$  & 1600
      &  4.887 &  4.887 & $\phantom{-}0.0$ & 1 & 1 & 0.18/0.18 & Tie \\
    \midrule
    \multicolumn{3}{l}{\textbf{Lower retained compliance under LLM policy: 4/6 \quad Ties: 2/6}} \\
    \multicolumn{3}{l}{Summed mean runtime over six problems (s):} & \multicolumn{2}{c}{6075 vs 2196} & $+177\%$ \\
    \bottomrule
  \end{tabular}
  \end{adjustbox}
\end{table}

\begin{figure}[!p]
  \centering
  \includegraphics[width=\linewidth]{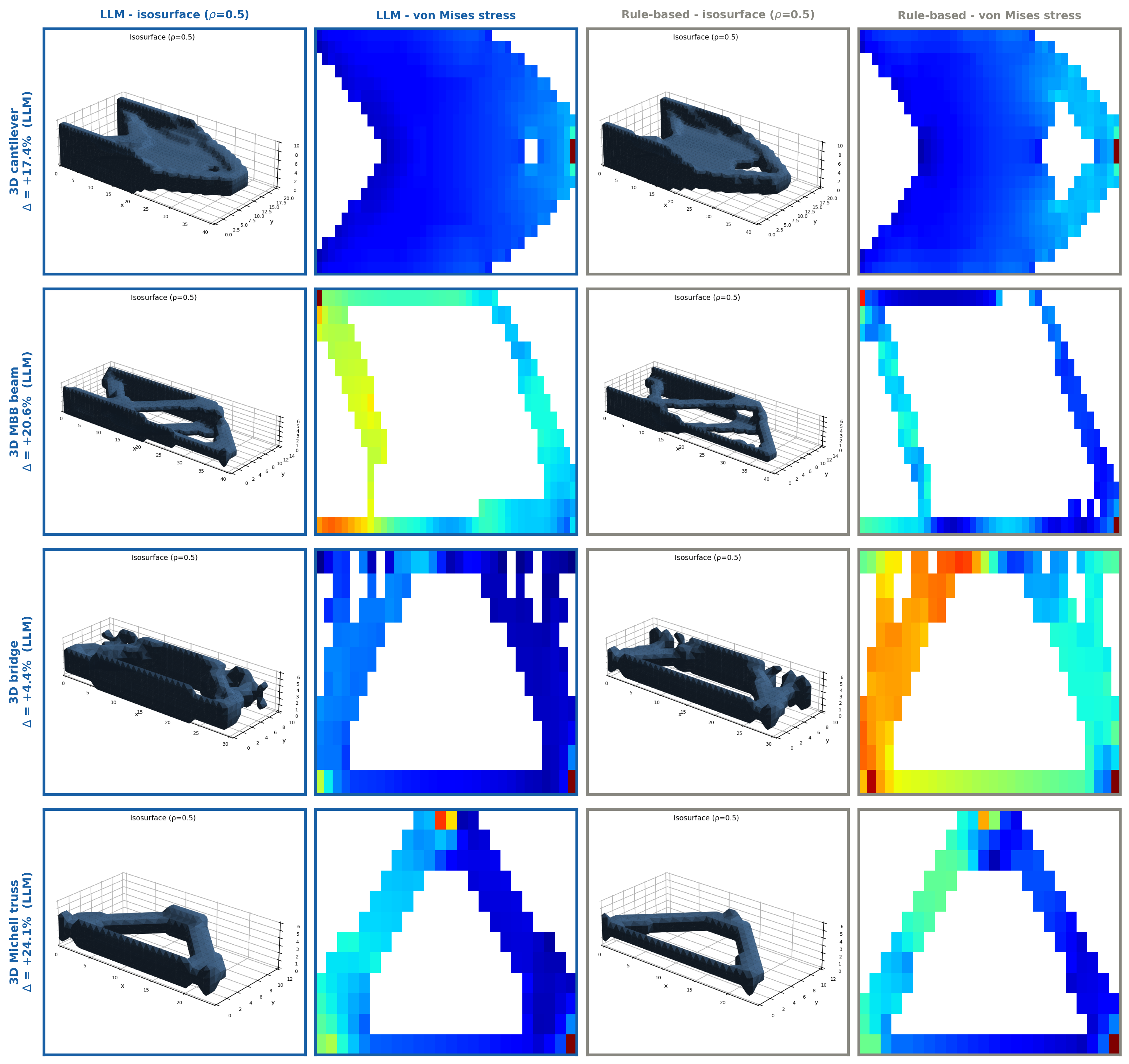}
  \caption{Isosurface renderings ($\rho_e = 0.5$ threshold) and
    von~Mises stress fields for the four non-tie 3D benchmark problems.
    Images are seed-42 examples; compliance percentages use three-seed mean
    retained-best compliance $\Crep$. Left two columns: LLM (blue border); right two columns:
rule-based (gray border). Stress colors are normalized per panel and
    are qualitative within-panel visualizations only; the 3D bridge LLM row
    uses the validity-filtered bridge evaluation checked in
    Appendix~\ref{app:repro_map}.}
  \label{fig:3d_gallery}
\end{figure}

\begin{figure}[!p]
  \centering
  \includegraphics[width=\linewidth]{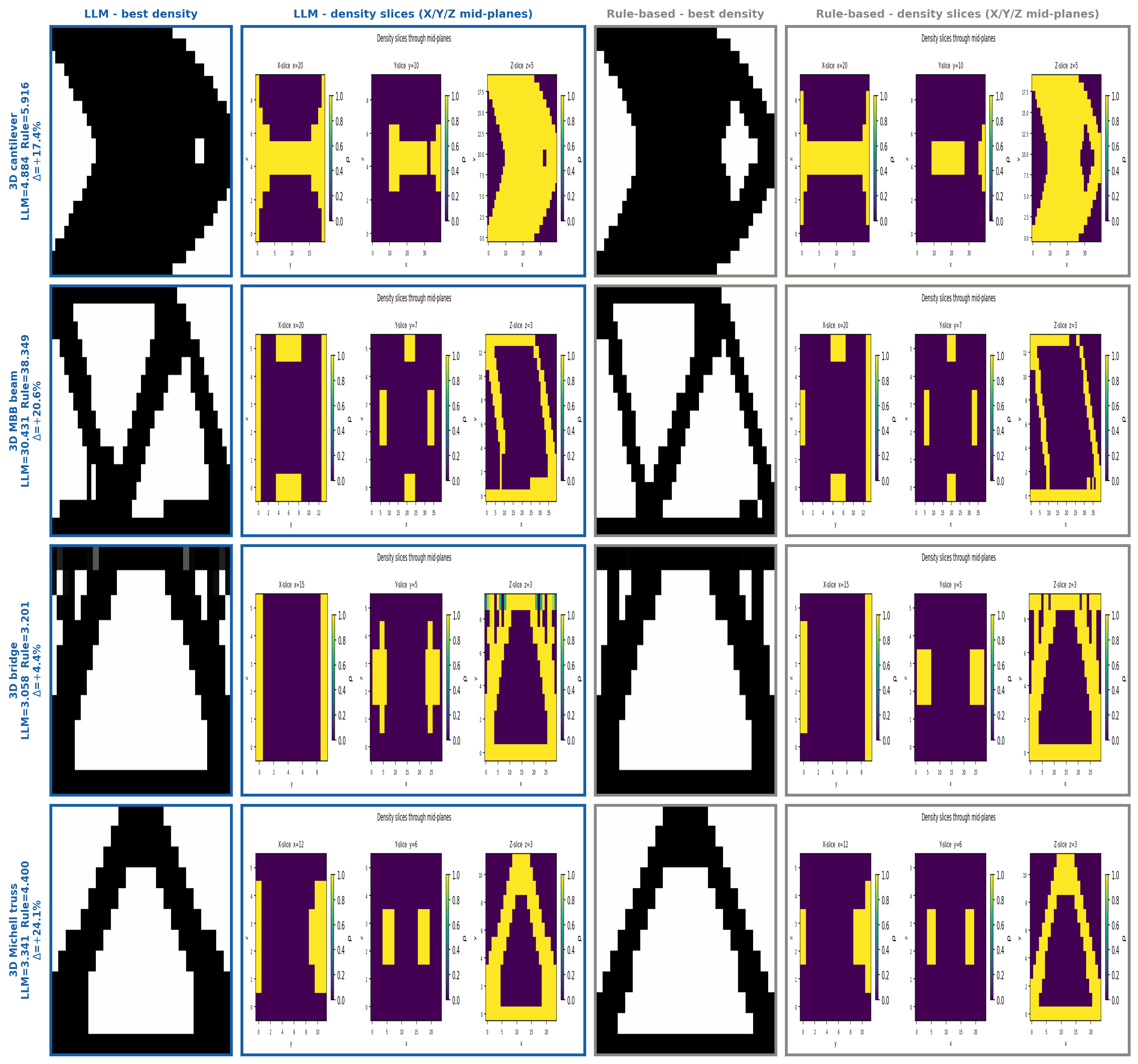}
  \caption{Representative seed-42 density fields and density slices through mid-planes
    (X, Y, Z) for the four non-tie 3D benchmark problems
    ($\sigyield = 120.3$).
    Scalar labels use three-seed mean retained-best compliance $\Crep$.
LLM (blue border) vs.\ rule-based (gray border). The 3D bridge LLM
    panels use the validity-filtered bridge evaluation passing the predefined gate checks
    in Appendix~\ref{app:repro_map}.}
  \label{fig:3d_slices}
\end{figure}

Figure~\ref{fig:3d_gallery} shows isosurface renderings and von~Mises
stress fields for the four non-tie 3D problems;
Figure~\ref{fig:3d_slices} provides matching density and density-slice
comparisons for the same four problems.

These 3D results should be interpreted with caution because the stress
gate is not separately calibrated for 3D and the comparison is not
fixed-volume.  \emph{Mechanistic limitation:} in the recorded 3D traces,
3D hotspot and member-widening actions add circular seed regions extruded
through depth rather than spherical volumetric seeds.  The LLM policy also
changes global parameters, including volume fraction, filter radius, and
penalization.  The retained-compliance differences in Table~\ref{tab:3d}
therefore cannot be attributed uniquely to seed materialization.
The mean 3D runtime across the reported runs is 6075~s for the LLM
controller and 2196~s for rule-based control, a $+177\%$ overhead.

\FloatBarrier

\section{Ablation Study}
\label{sec:ablation}

\subsection{Three-condition comparison}

\begin{figure}[!htbp]
  \centering
  \includegraphics[width=\linewidth]{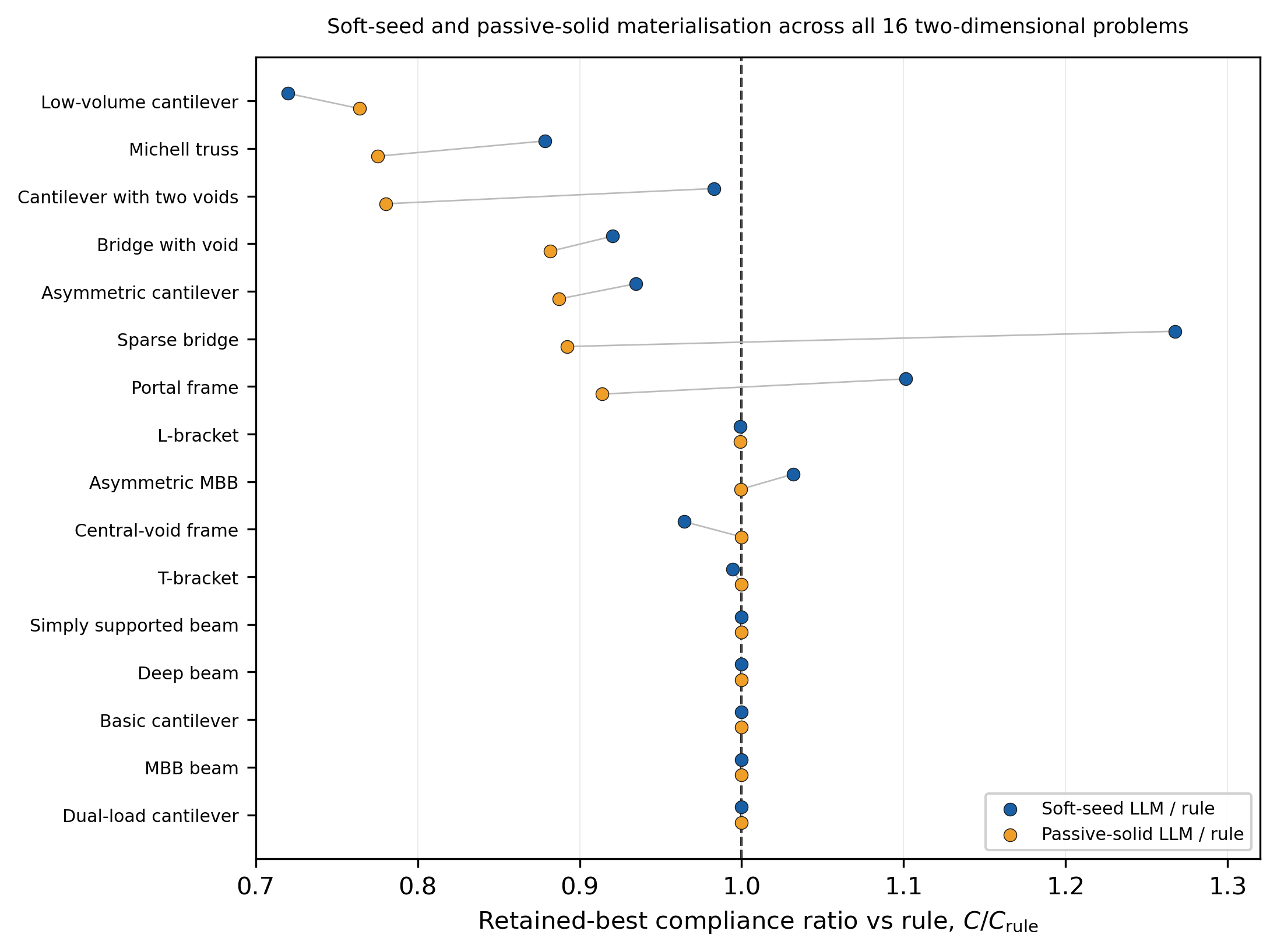}
  \caption{Three-way ablation summarized as retained-compliance ratios for all
    16 two-dimensional problems.  Blue points show soft-seed LLM divided by
    rule-based retained-best compliance; amber points show passive-solid LLM
    divided by rule-based retained-best compliance.  Values below one are lower
    than the rule-based controller.  The plot makes visible that passive-solid
materialization is the stronger aggregate condition, while soft seeding is
    the less restrictive mechanism.}
  \label{fig:ablation}
\end{figure}

Figure~\ref{fig:ablation} summarizes all 16 two-dimensional problems, while
Table~\ref{tab:ablation} reports representative cases where the three
conditions produce meaningfully different outcomes.  The values are three-seed
means from the reported ablation summaries.

\begin{table}[!htbp]
  \caption{Three-way ablation at $\sigyield = 120.3$.
    Soft-seed LLM: soft density seeding.
    Passive-solid LLM: passive-solid seeding.
    Rule: rule-based.
    $\Delta$ columns: relative retained-compliance differences; positive
    values denote lower retained compliance for the left named condition.}
  \label{tab:ablation}
  \centering
  \small
  \begin{tabular}{p{3.4cm}rrrrrr}
    \toprule
    Problem & Soft LLM & Solid LLM & Rule
      & $\Delta$ soft vs rule & $\Delta$ soft vs solid \\
    \midrule
    Low-volume cantilever      &  99.7 & 105.8 & 138.5 & $+28.0\%$ & $+5.8\%$ \\
    Michell truss         &  26.0 &  23.0 &  29.6 & $+12.1\%$ & $-13.3\%^*$ \\
    Central-void frame       &  17.6 &  18.3 &  18.3 & $+3.5\%$  & $+3.5\%$ \\
    Cantilever with two voids & 115.2 &  91.5 & 117.2 & $+1.7\%$  & $-25.9\%^*$ \\
    Portal frame          &  89.8 &  74.5 &  81.5 & $-10.1\%$ & $-20.5\%^*$ \\
    Sparse bridge         & 371.9 & 261.8 & 293.4 & $-26.8\%$ & $-42.1\%^*$ \\
    \midrule
    \multicolumn{3}{l}{Geo.\ mean, all 16 problems:} & & & \\
    \quad $C_{\text{soft}}/C_{\text{rule}}$ & & & $= 0.981$ & $(+1.9\%)$ & \\
    \quad $C_{\text{solid}}/C_{\text{rule}}$ & & & $= 0.926$ & $(+7.4\%)$ & \\
    \quad $C_{\text{soft}}/C_{\text{solid}}$  & & &           & & $= 1.059\ (-5.9\%)$ \\
    \bottomrule
  \end{tabular}
  \medskip

  \noindent
  \small $^*$ Passive-solid seeding outperforms soft seeding on these
  problems.  This is the main reason passive-solid intervention is the
  stronger aggregate condition in the reported data.
\end{table}

\FloatBarrier

\subsection{Ablation interpretation}

Across all 16 problems, the geometric mean compliance ratios are:
\begin{equation}
  \begin{aligned}
  C_{\text{soft}} / C_{\text{rule}} &= 0.981,\\
  C_{\text{solid}} / C_{\text{rule}} &= 0.926,\\
  C_{\text{soft}} / C_{\text{solid}} &= 1.059.
  \end{aligned}
\end{equation}

These ratios do not support a positive additive decomposition in which
soft seeding improves on passive-solid materialization.  Instead,
passive-solid intervention is the stronger aggregate condition in the
reported data, while soft seeding remains mechanistically attractive because
it does not remove optimizer freedom.  Thus, in this benchmark, soft seeding
is the less restrictive but weaker materialization mechanism, not the
strongest controller-policy variant.

Per-problem behavior is mixed.  Soft seeding is better than passive-solid
materialization on the low-volume cantilever and central-void frame, but
passive-solid materialization is better on the Michell truss,
cantilever with two voids, portal frame, and sparse bridge.  These cases
motivate future work on adaptive seed persistence rather than a fixed
choice between soft and frozen materialization.

\FloatBarrier

\section{Discussion}
\label{sec:discussion}

\subsection{Stress constraint handling: outer-loop gating vs.\ gradient-based formulations}

A fundamental design choice in \methodname{} is that stress constraints
are handled via an \emph{outer design loop} rather than within the SIMP
solver's gradient computation.  At each design step, the inner solver
minimizes compliance only; a separate forward FEA pass then evaluates the
per-element von~Mises stress field, and the evaluator checks whether
$\max_{e:\rho_e > 0.5} \sigvm^{(e)} \leq \sigyield$.  If this gate
fails, the LLM proposes a spatially targeted geometry modification and the
solver is re-invoked on the updated specification.

This is a deliberate architectural choice, not an equivalence claim, and it has
concrete consequences.  The SIMP density update receives no gradient
information about the stress field: sensitivities $\partial \sigvm / \partial \rho_e$
are never computed, and the optimality-criteria update (Equation~\eqref{eq:oc_update})
drives densities toward compliance reduction without direct knowledge of where
stress concentrations are forming.  In classical stress-constrained
formulations---P-norm penalty~\cite{le2010stress,yang1996stress}, augmented
Lagrangian~\cite{senhora2020topology}, or $\varepsilon$-relaxation~\cite{duysinx1998topology}---the
stress constraint enters the Lagrangian directly, and adjoint analysis
provides element-wise stress sensitivity to guide the density update.
These methods can resolve stress-feasible topologies in a single solve
without an outer loop.

The rationale for our outer-loop approach is threefold.  First, it
preserves the unmodified SIMP solver as a black box, enabling the
framework to wrap any existing density-based solver without modification
to its adjoint computation or sensitivity filtering.  Second, the LLM's
contribution is a form of visual, geometric intervention that is
complementary to adjoint stress sensitivities: it identifies \emph{where} a
stress concentration appears in a rendered field and proposes a geometric
fix at that location.  Adjoint stress sensitivities tell the solver
\emph{how much} each element contributes to an aggregate stress measure;
the visual controller supplies an interpretable spatial action vocabulary.
Third, the outer-loop
formulation naturally accommodates the soft density seed mechanism:
geometry modifications are applied to the problem specification between
solver calls, and the seed initialization biases the next solve without
constraining it.

The limitation is equally clear: when the stress constraint is tight
(Section~\ref{subsec:opwindow}), the five-step outer loop may be
insufficient to drive peak stress below $\sigyield$ through geometry
modifications alone, whereas a gradient-based formulation would
continuously adjust densities toward stress feasibility.  The deterministic
calibration study (Table~\ref{tab:sensitivity}) is consistent with this limitation: at $q=40$
($\sigyield = 110.4$), both conditions fail on 6 of 16 problems because
the compliance-only topology violates the stress gate by too wide a margin
for spatial seeding to resolve within the step budget.

A natural extension---combining the LLM's spatial targeting with a
gradient-based stress-constrained inner solver---would address this
limitation by providing adjoint stress sensitivity within each solve while
using the LLM to propose inter-solve geometry modifications for
concentrations that persist across continuation steps.  We leave this
integration to future work.

\subsection{Operating window}
\label{subsec:opwindow}

The deterministic calibration study identifies a moderate stress-gate range
($\sigyield \in [120, 149]$, calibration percentile $q\in [50, 55]$) in which
roughly half the 2D problems fail the compliance-only stress gate.  Below
this range, constraints are tight enough that neither spatial seeding nor
volume-fraction adjustments can satisfy the hardest cases within five
design steps.  Above it, the gate becomes loose and additional spatial
reasoning is less likely to be repaid.

This operating window is consistent with the architectural role of an
outer-loop spatial meta-heuristic: it is most relevant when the
compliance-only topology is close to stress-feasible but requires localized
geometric corrections.

\subsection{Mesh resolution and scalability}

The benchmark problems in this study use relatively coarse meshes
(1,600--8,000 elements), which are sufficient for an exploratory evaluation
of the LLM spatial targeting mechanism but below the resolution of
production-scale stress-constrained topology optimization studies that routinely employ
10,000--100,000
elements~\cite{senhora2020topology,holmberg2013stress,dasilva2021threedim}.
At coarse resolution, stress concentrations are partially smoothed by the
mesh itself, potentially understating the difficulty of the stress
constraint.  The current 3D cases are therefore exploratory small-mesh tests
and do not establish scalability.

The scalability of \methodname{} is governed by two factors: the SIMP
solver's FEA cost, which grows with the number of elements $n_e$, and
the LLM call cost.  The interpreter receives fixed-resolution
$300{\times}300$ images regardless of the underlying mesh, so API latency is
less directly tied to mesh size than the FEA cost.  The LLM overhead would
therefore be expected to become a smaller fraction of total wall time if FEA
cost dominates at higher mesh resolution.
Empirically, the 3D benchmark (up to 8,000 elements) shows lower retained
compliance for the LLM controller-policy condition on 4 of 6 problems and ties on 2,
but this should not be read as an isolated validation of 3D seed
materialization because no matched no-seed 3D ablation is included.
Validation at higher resolutions and with a separately calibrated 3D stress
gate remains an important direction for future work.

\subsection{Computational overhead}

The soft-seed LLM condition incurs an $+88\%$ wall-time overhead over
rule-based control on the 16 two-dimensional problems (about 11.4~min vs
6.0~min in summed mean runtime on CPU).
The additional wall time is consistent with one hosted-model call per design
step ($\approx 5$--$10$~s in observed interactive use), but per-call provider
latency was not persisted in the traces. The additional FEA pass for stress
computation adds approximately one solver iteration equivalent.
The reported primary comparison comprises 66 LLM-controller runs and
66 matched rule-based runs (48 2D and 18 3D runs per condition);
the 3D bridge LLM value uses the validity-filtered bridge evaluation checked
in Appendix~\ref{app:repro_map}.
Trace and action counts are reported in Appendix~\ref{app:repro_map}.  The
summed recorded wall time is
20,272~s (5.63~h) for the LLM traces and 7,676~s (2.13~h) for the
rule-based traces on the CPU-only Windows workstation documented in the
computational-environment notes and summarized in Appendix~\ref{app:repro_map}.  Exact model-service token
counts, per-call latency distribution, energy use, retention settings, and
served model revision are not available from the archived records, so exact
hosted-model replay is not possible.
This overhead is material and should be reduced by batching API calls,
skipping interpreter calls once gates pass, or using locally hosted models.

\subsection{Failure modes and future directions}

Four failure patterns were identified: high seed sensitivity on some 2D
problems, geometric imprecision in seed placement, soft-seed erosion when
a persistent reinforcement would be beneficial, and stress saturation under
very tight constraints ($q\leq 45$), where the compliance-only topology
violates the stress gate by too wide a margin for five design steps to
resolve.

The fixed-volume 2D study controls the largest material-budget confound and
adds direct localization measurements, but it does not close the causal
question.  Feasible-final scoring is mixed, deterministic hotspot seeding is
competitive, several policies improve retained-any compliance without a
matching feasible-primary advantage, and the LLM condition has incomplete
execution/trace coverage.  The next evidence tier is therefore not additional
prompt tuning; it is comparison against a stress-aware inner solver such as
P-norm, K--S, or augmented Lagrangian stress optimization, followed by
separately calibrated fixed-volume 3D reruns and mesh-refinement checks.

\subsection{Relationship to prior work}

Compared to the compliance-only predecessors
\cite{yang2026itersimp,yang2026autosimp}, which operated on
a single density image and used only passive void/solid regions, the
stress-gated extension introduces two key advances: the dual-image
interpreter that conditions on both density and stress fields simultaneously,
and the soft density seed mechanism that preserves optimizer freedom
(Section~\ref{subsec:3d} discusses the 3D attribution limits).
More specifically, the earlier adaptive-continuation work studied
compliance-only schedule control, while AutoSiMP studied natural-language
problem configuration and end-to-end setup; neither evaluates dual
density--stress visual inputs, local stress-violation localization,
stress-gated spatial actions, or the fixed-volume attribution tests reported
here.
The ablation in Section~\ref{sec:ablation} shows that the materialization
choice is unresolved: passive-solid intervention is stronger in aggregate,
while soft seeding preserves optimizer freedom and remains the less
restrictive mechanism for future stress-aware inner solvers.

\section{Conclusion}
\label{sec:conclusion}

We have presented \methodname{}, an exploratory framework for stress-gated
topology optimization in which an LLM agent interprets dual
density-and-stress field images and proposes spatially targeted geometry
modifications via an outer design loop that wraps an unmodified
compliance-only SIMP solver.  The current evidence supports a narrower claim
than a general performance-method claim: LLM spatial proposals are technically
implementable and inspectable, with selected cases showing benefit, but the
fixed-volume study does not establish visual reasoning as the causal source of
performance improvement.

The key results are:
(i)~the soft-seed LLM condition gives a modest 1.9\% lower geometric-mean
retained compliance than rule-based control on the 16 2D problems, under
controller policies that can change final volume fraction, but the two-sided
Wilcoxon test is not significant ($W=33$, $p=0.382$; one-sided $p=0.191$);
when the recorded traces are re-scored under feasible-retained and
feasible-final metrics, infeasible runs are excluded rather than counted as
wins, leaving five and six incomplete 2D pairs respectively;
(ii)~the passive-solid LLM condition gives a larger 7.4\% lower
geometric-mean retained compliance than rule-based control, but it achieves
this by freezing LLM-placed material and is therefore a more restrictive
mechanism than soft seeding;
(iii)~the fixed-volume 2D study gives lower LLM feasible-retained compliance
than rule-based control in 5 of 9 complete paired problems; 44 of 48 LLM attempted slots
completed a design evaluation, 25 of those 44 completed evaluations have an
all-gate-passing retained state, and 27 of 48 rule-based evaluations have an
all-gate-passing retained state,
feasible-final scoring is split 4/4/1, and deterministic max-stress hotspot
seeding remains competitive;
(iv)~direct localization over 76 accepted LLM spatial actions with per-step records gives mean
normalized seed-to-hotspot distance 0.221 and mean overlap with the top
1\%, 5\%, and 10\% stress regions of 0.041, 0.118, and 0.196, so the current
evidence does not prove precise visual hotspot localization;
(v)~several soft-seed cases have lower retained compliance, most notably the
low-volume cantilever ($+28.0\%$), Michell truss ($+12.1\%$), bridge with
circular void ($+8.0\%$), and asymmetric low-volume cantilever ($+6.5\%$), but
these are retained-best controller-policy differences rather than
feasible-final fixed-volume wins;
(vi)~the six-problem 3D study remains exploratory controller-level evidence,
not a fixed-volume or separately calibrated 3D validation; and
(vii)~the 2D computational overhead is $+88\%$ in wall time and the 3D subset
has about $+177\%$ summed runtime overhead, so the current system should be
read as an inspectable design-assist and spatial-proposal framework rather than
an efficient optimizer, although per-call provider latency was not persisted.

The stress constraint in \methodname{} is handled as an outer-loop gate
check rather than through gradient-based constraint handling within the
solver. This preserves solver modularity and motivates further study of whether
LLM-guided spatial interventions can complement established adjoint-based
methods.
Integrating LLM spatial targeting with gradient-based stress-constrained
formulations is a natural next step that could combine the spatial-action
traces studied here with the continuous sensitivity guidance that
adjoint methods provide.

\FloatBarrier

\section*{Code availability}

The full code release will be made available after the arXiv preprint is online
at the
\href{https://github.com/nbbllxx0/IterSIMP-sigma-LLM-Assisted-Spatial-Interventions-for-Stress-Aware-Topology-Optimization}{project GitHub repository}.
It will include runnable code, environment files, experiment drivers,
figure-generation scripts, smoke tests, compact result summaries, and a
comprehensive README for reproducing the reported and extended experiments.
No external dataset is used. The release is intended for code-level
reproduction and new reruns; it will include the scripts and compact result
summaries needed to reproduce the reported figures and tables. It will also
include prompt templates, action-validation logic, and the redacted audit-record
schema used by new API-backed reruns to retain rendered inputs, provider
outputs, parsed actions, and solver-state summaries. Exact replay of the
original hosted-model calls is not available because complete provider I/O
records, response identifiers, served-model revision metadata, and retention
settings were not archived.

\section*{Ethics and data-transfer note}

The experiments use synthetic structural benchmark fields and no human-subject
or personal data. LLM-controlled runs send rendered density and stress images
plus accompanying text and numerical prompts to a hosted model; proprietary applications
should use available data-retention controls or a local/on-premise vision model.
Provider energy use was not observable, so the paper reports local wall-clock
timing only. Controller outputs are research evaluations and require
conventional engineering verification before safety-critical design use.

\section*{Funding}

This research was supported by startup funding from Santa Clara University.

\section*{Declaration of competing interest}

The authors declare that they have no known competing financial interests or
personal relationships that could have appeared to influence the work
reported in this paper.

\appendix
\section{Reproducibility Map and Metric Definitions}
\label{app:repro_map}

Table~\ref{tab:repro_map} maps the manuscript-level claims to the evidence
used to generate them.  The scalar tables use $\Crep$ as
defined in Section~\ref{sec:experiments}; this is a retained-best diagnostic
metric, not a feasibility-conditioned primary performance metric. $\Cfinal$ is
retained in the per-step records and is reported only in explicitly labeled
feasible-final re-scores and fixed-volume endpoints.

\begin{table}[htbp]
  \caption{Compact reproducibility map for the reported claims.}
  \label{tab:repro_map}
  \centering
  \small
  \begin{tabularx}{\linewidth}{p{4.0cm}Y}
    \toprule
    Paper element & Evidence used for reporting \\
    \midrule
    Main 2D and exploratory 3D scalar tables &
    Retained-best compliance summary with final gate-pass counts reported
    separately. The 3D bridge row uses a validity-filtered bridge evaluation. \\
    Feasibility-conditioned re-score &
    Existing-trace feasibility table; infeasible traces are excluded from
    feasible-retained and feasible-final means. \\
    Fixed-volume 2D attribution study &
    Fixed-volume summary and tables; endpoint tables
    separate attempted slots, completed evaluations, missing executions, and
    feasible completed evaluations. Recovered scalar records without per-step
    traces are excluded from localization diagnostics. \\
    Sensitivity and ablation analyses &
    Sensitivity, Wilcoxon, and mechanism-ablation summaries. \\
    Figure panels &
    Figure assets generated from the reported summaries and retained trace
    images; representative galleries use seed-42 examples unless
    stated otherwise. \\
    Model/runtime configuration &
    Gemini 3.1 Flash-Lite Preview at temperature 0, seeds 42/123/7, Python
    3.11.5, NumPy 1.24.3, SciPy 1.11.1, Matplotlib 3.7.2, Pillow 9.4.0,
    Windows 10.0.26200, Intel Core i9-13900KF, 32 logical processors, and
    95.85 GiB visible memory. \\
    Hosted-model replay scope &
    The archived records retain parsed prompts/action summaries and solver
    state, but not complete provider input/output records, provider response
    identifiers, served-model revision metadata, or retention settings. \\
    \bottomrule
  \end{tabularx}
\end{table}

The reported 3D bridge LLM row uses a validity-filtered bridge evaluation for the same benchmark setting that passes
the predefined finite-positive-compliance validity criterion. An earlier 3D
bridge run had a nonphysical retained-compliance value and is excluded from
the scalar table and visual panels.

\begin{table}[htbp]
  \caption{Reader-facing metric definitions used in the paper.}
  \label{tab:metric_fields_compact}
  \centering
  \small
  \begin{tabularx}{\linewidth}{p{3.6cm}Y}
    \toprule
    Metric & Meaning \\
    \midrule
    Retained-best compliance, $\Crep$ &
    Retained-best compliance across design steps; this is the scalar used in
    the compliance tables and geometric means and is not necessarily an
    all-check-passing state. \\
    Final-state compliance, $\Cfinal$ &
    Compliance of the last solver state in a trace; available for inspection but not
    used in the scalar tables unless explicitly stated. \\
    Gate-pass status &
    The final state passes all seven gate checks. Four informational
    diagnostics are recorded separately and do not gate validity. \\
    Population standard deviation &
    Descriptive standard deviation over the three repeats; it is not an
    inferential uncertainty estimate. \\
    \bottomrule
  \end{tabularx}
\end{table}

\begin{table}[htbp]
  \centering
  \caption{Evaluation availability and controller-action counts for fixed-volume
  localization diagnostics.}
  \label{tab:trace_action_counts}
  \scriptsize
  \begin{tabular}{lrrr}
    \toprule
    Condition & Evaluations with records & Accepted controller actions & Rejected or inadmissible actions \\
    \midrule
    Soft-seed LLM & 28 & 84 & 0 \\
    Exact hotspot & 47 & 148 & 0 \\
    Rule-based controller & 48 & 78 & 0 \\
    Random stress region & 48 & 155 & 0 \\
    Density-only LLM & 48 & 141 & 0 \\
    Stress-only LLM & 48 & 150 & 0 \\
    Numeric-only LLM & 48 & 144 & 0 \\
    Global-only LLM & 48 & 81 & 0 \\
    \bottomrule
  \end{tabular}

  \medskip
  \parbox{\linewidth}{\scriptsize Accepted controller actions include all
  accepted actions recorded in per-step records. Localization statistics in the
  main text and figures use the accepted spatial-action subset; for the
  soft-seed LLM condition this subset contains 76 spatial actions over
  28 completed evaluations with per-step records.}
\end{table}

\begin{table}[htbp]
  \caption{Initial and final volume-fraction targets for the 2D primary
  controller-policy comparison. Values are means over three traces; parentheses
  list distinct final targets when repeats differ.}
  \label{tab:final_vf_2d}
  \centering
  \scriptsize
  \begin{tabular}{lccc}
    \toprule
    Problem & Initial $v_f$ & Soft-seed LLM final $v_f$ & Rule-based final $v_f$ \\
    \midrule
    Asymmetric low-volume cantilever & 0.18 & 0.56 & 0.50 \\
    Sparse bridge & 0.18 & 0.56 & 0.50 \\
    Bridge with circular void & 0.25 & 0.61 (0.57/0.63) & 0.57 \\
    Low-volume cantilever & 0.12 & 0.33 (0.29/0.35) & 0.28 \\
    Basic cantilever & 0.35 & 0.35 & 0.35 \\
    Cantilever with two voids & 0.25 & 0.28 & 0.28 \\
    Deep beam & 0.35 & 0.35 & 0.35 \\
    Dual-load cantilever & 0.30 & 0.46 & 0.46 \\
    Central-void frame & 0.22 & 0.18 & 0.12 \\
    L-bracket & 0.30 & 0.30 & 0.30 \\
    Asymmetric MBB beam & 0.30 & 0.59 (0.54/0.62) & 0.62 \\
    MBB beam & 0.35 & 0.67 & 0.67 \\
    Michell truss & 0.15 & 0.16 (0.08/0.20) & 0.08 \\
    Portal frame & 0.22 & 0.25 (0.18/0.22/0.36) & 0.30 \\
    Simply supported beam & 0.30 & 0.30 & 0.30 \\
    T-bracket & 0.25 & 0.17 (0.15/0.21) & 0.15 \\
    \bottomrule
  \end{tabular}
\end{table}

Regeneration commands and checksums are provided with the released code rather
than duplicated in the article body.

\FloatBarrier
\bibliographystyle{unsrtnat}
\bibliography{referencesL4}

\end{document}